\documentclass[iop,apj]{emulateapj}

\begin{document}

\title{Massive Galaxies are Larger in Dense Environments: Environmental Dependence of Mass--Size Relation of Early-Type Galaxies}
\shorttitle{Environmental Dependence of Mass--Size Relation of Early-Type Galaxies}
\shortauthors{Yoon et al.}

\author{Yongmin Yoon\altaffilmark{1}, Myungshin Im\altaffilmark{1}, and Jae-Woo Kim\altaffilmark{1,2}
}

\email{yymx2@astro.snu.ac.kr, mim@astro.snu.ac.kr}

\altaffiltext{1}{Center for the Exploration of the Origin of the Universe (CEOU),
Astronomy Program, Department of Physics and Astronomy, Seoul National University, 599 Gwanak-ro, Gwanak-gu, Seoul, 151-742, Korea} 
\altaffiltext{2}{Korea Astronomy and Space Science Institute, 776 Daedeokdae-ro, Yuseong-gu, Daejeon, 34055, Republic of Korea}

\begin{abstract}
 Under the $\Lambda$ cold dark matter ($\Lambda$CDM) cosmological models, massive galaxies are expected to be larger in denser environments through frequent hierarchical mergers with other galaxies. Yet, observational studies of low-redshift early-type galaxies have shown no such trend, standing as a puzzle to solve during the past decade. We analyzed 73,116 early-type galaxies at $0.1\leq z < 0.15$, adopting a robust nonparametric size measurement technique and extending the analysis to many massive galaxies. We find for the first time that local early-type galaxies heavier than $10^{11.2}M_{\odot}$ show a clear environmental dependence in mass--size relation, in such a way that galaxies are as much as 20 -- 40 \% larger in densest environments than in underdense environments. Splitting the sample into the brightest cluster galaxies (BCGs) and non-BCGs does not affect the result. This result agrees with the $\Lambda$CDM cosmological simulations and suggests that mergers played a significant role in the growth of massive galaxies in dense environments as expected in theory.
\\
\end{abstract}

\keywords{galaxies: elliptical and lenticular, cD --- galaxies: evolution --- galaxies: formation ---  galaxies: statistics --- galaxies: structure}

\section{Introduction} \label{sec:Intro}
 Early-type galaxies such as ellipticals and bulge-dominated lenticular galaxies are expected to form and grow through a number of major and minor mergers in the $\Lambda$ cold dark matter ($\Lambda$CDM) universe \citep{Baugh1996,Christlein2004,vanDokkum2005,DeLucia2006,Khochfar2006,DeLucia2007,Wilman2013}. Today, the most massive galaxies in the universe are mostly bulge-dominated early-type galaxies, which is consistent with this view of the galaxy evolution. If the hierarchical galaxy formation model in the $\Lambda$CDM universe is correct, these massive early-type galaxies should reveal properties that reflect the model predictions.

 Sizes of galaxies have been recognized as one of the key galaxy properties that are useful for testing the galaxy evolution models \citep{Mutz1994, Casertano1995, Im1995, Fan2008, Bezanson2009, Damjanov2009, Naab2009, vanderWel2009, Fan2010, Hopkins2010, vanDokkum2010,  R&G2011, Carollo2013, Ishibashi2013, Oogi2013, Shankar2013}. For early-type galaxies, sizes can reflect their merger history -- dissipationless mergers such as dry mergers can increase galaxy sizes in proportion to mass change, while dissipational merger (wet merger) does not, leading to a scale relation that is different at low and high mass. Indeed, many studies have shown that the mass--size relation of early-type galaxies can be a result of hierarchical growth \citep{Shen2003,Bernardi2007,vandenBergh2008,Hyde2009,Maltby2010,Nair2010,Valentinuzzi2010,Bernardi2011b,Huertas2013b}. It is found that the most massive part ($M_{\star}>2\times10^{11}\,M_{\odot}$) of the mass--size relation is curved in such a way that their sizes are larger than the linear relation of less massive galaxies \citep{Desroches2007,Hyde2009,Bernardi2011b}. The curvature of scaling relations in early-type galaxies can be found in other relations such as color--, velocity dispersion--, axis ratio--, and color gradient--mass \citep{Bernardi2011a,Bernardi2011b}. Interestingly, all the relations have a special point at $M_{\star}\sim2\times10^{11}\,M_{\odot}$ where the slopes begin to change. The studies suggest that this is due to an increased role of galaxy assembly by major mergers above this point. 

 The mass--size relation can be examined further as a function of environment to test the hierarchical galaxy formation models. The $\Lambda$CDM cosmological models predict that galaxies form early in dense environments, and they are expected to undergo more merging activities than in less dense environments. If early-type galaxies are products of merging activities, their properties are expected to reflect the environmental dependence in the $\Lambda$CDM cosmological models. For example, their color gradients are found to be flatter in cluster environments due to mixing of stars through frequent merging \citep{Ko2005,LaBarbera2005,Tortora2012}.  

 The past merger history should be reflected in sizes of early-type galaxies as well. In terms of environmental dependence, we expect the following. The galaxy evolution is accelerated in dense environments, so that massive galaxies formed early and had undergone merging activities and their sizes grew rapidly through dry merging at the massive end. On the other hand, early-type galaxies in less dense environments are catching up the evolutionary stages that cluster early-type galaxies underwent, and their sizes are relatively smaller for a given galaxy mass.  

 Consequently, hierarchical galaxy simulation models predict the environmental dependence of the mass--size relation \citep{Shankar2013,Shankar2014}. These models predict that galaxies living in massive halos today are larger than their counterparts by a factor of $\sim1.5$ -- $3.0$.
  
 In spite of such a prediction, previous studies on early-type galaxies at $z\lesssim0.2$ have found that there is no environmental dependence in their mass--size relation \citep{Maltby2010,Nair2010,Cappellari2013,Huertas2013b}. On the other hand, there are studies suggesting that early-type galaxies in cluster environments are larger ($\sim30$ -- $50\%$ or more) than their counterparts in fields at $1\lesssim z\lesssim2$ \citep{Papovich2012,Bassett2013,Lani2013,Strazzullo2013,Delaye2014}. In contrast, \citet{Rettura2010} could not find environmental dependence at the high redshift. At intermediate redshift ($0.2 < z < 1.0$), studies show mixed results with \citep{Cooper2012} or without \citep{Huertas2013a,Kelkar2015} environmental dependence. Some studies even claim that field galaxies are even slightly larger than their counterparts in clusters \citep{Raichoor2012,Poggianti2013,Cebrian2014}. To sum up, the observational results at low redshift disagree with the theoretical predictions and the results for $1\lesssim z\lesssim2$. Results at intermediate redshifts are mixed. The current status of the observational studies is summarized in Table \ref{tabint} with various information and special remarks.

 The discrepancy between the low-redshift results and the model predictions can be possibly caused by the mass range considered. For example, the mass range of the sample of \citet{Cappellari2013} is limited to $\log M_{\star} \la 11.5$, and they have a small number of objects ($\sim10$) at $\log (M_{\star}/M_{\odot})>11.3$, making it difficult to find any environmental dependence beyond $\log (M_{\star}/M_{\odot})>11.2$. The small sample sizes of some of the studies, especially at the massive end, also make it difficult to judge the robustness of the results. Additionally, a decrease of the environmental dependence from high redshift to low redshift \citep{Lani2013,Delaye2014} could be the reason for the contradiction between results for high redshift and low redshift.

 In this paper, we provide enough number statistics especially at massive end in the study of the mass--size relation. More specifically, we examine environmental dependence of the mass--size relation of 73,116 early-type galaxies in the local universe ($0.1\lesssim z<0.15$), especially for massive galaxies ($10.7\lesssim \log (M_{\star}/M_{\odot})\lesssim11.9$). They are on top of the  hierarchical assembly tree; therefore, we can expect that they have had more chances to be affected by environment in their assembly histories. Through this work, we report that environmental dependence exists in the mass--size relation of the most massive early-type galaxies at low redshift with $\log (M_{\star}/M_{\odot})\ga11.2$, in qualitative agreement with the predictions of semianalytical models of hierarchical galaxy formation.

Throughout this paper, we use \emph{H$_0=70$} km s$^{-1}$Mpc$^{-1}$, $\Omega_{\Lambda}=0.7$, and $\Omega_{m}=0.3$ as cosmological parameters \citep[e.g.,][]{Im1997} and adopt the AB magnitude system.
\\

\begin{deluxetable*}{lccccc}
\tablecaption{Studies for Mass--Size Relation of Early-Type Galaxies} 
\tabletypesize{\scriptsize}
\tablehead{
\colhead{Reference} & \colhead{Redshift} & \colhead{Mass Range\tablenotemark{s}}  & \colhead{Sample} & \colhead{Size} & \colhead{Environmental} \\
\colhead{} & \colhead{Range} & \colhead{$[\log (M_{\star}/M_{\odot})]$} & \colhead{Size\tablenotemark{t}} & \colhead{Measurement} & \colhead{Dependence}
}
\startdata 
Low Redshift, $z<0.2$\\

\citet{Maltby2010}\tablenotemark{a}& $z\sim0.167$ & $9.0\la \log M_{\star} \la 11.5$ & $\sim500$ & Single S\'{e}rsic & No\\

\citet{Nair2010}\tablenotemark{b}& $0.01<z<0.1$ & $9.6\la \log M_{\star} \la 11.6$ & $\sim5000$ & Petrosian $R_{90}$ & No \\

\citet{Cappellari2013}\tablenotemark{c}& $z\la0.02$ & $10.0\la \log M_{\star} \la 11.5$ & $\sim300$ & Growth Curve & No \\

\citet{Huertas2013b}\tablenotemark{d}& $z<0.09$ & $10.5\la \log M_{\star} \la 11.8$ & $\sim12,000$ & Single S\'{e}rsic & No\\

\citet{Poggianti2013}\tablenotemark{e} &  $0.03\le z\le0.11$ & $10.5\la \log M_{\star} \la 11.7$ & $\sim1500$ & Single S\'{e}rsic & Yes (Opposite)\\

\citet{Cebrian2014}\tablenotemark{f}& $0.005<z<0.12$ & $10.0\la \log M_{\star} \la 11.3$ & $\sim 10^5$ & Single S\'{e}rsic & Yes (Opposite) \\

\citet{Zhao2015}\tablenotemark{g}(BCG)& $0.02\le z\le0.10$ & $10.8\la \log M_{\star} \la 11.8$ & 425 & Single S\'{e}rsic & No\\

\textbf{This study}\tablenotemark{h} & $\textbf{0.1}\le \textbf{\textit{z}} < \textbf{0.15}$ &  $\textbf{10.7}\la \log M_{\star} \la \textbf{11.9}$ & \textbf{73,116} & \textbf{Growth Curve} & \textbf{Yes} \\

\tableline
Intermediate Redshift, $0.2\la z \la1$\\

\citet{Cooper2012}\tablenotemark{i}& $0.4<z<1.2$ & $10.0 \la \log M_{\star} \la 11.0$ & $\sim600$ & Single S\'{e}rsic & Yes\\

\citet{Huertas2013a}\tablenotemark{j}& $0.2<z<1.0$ & $10.5\la \log M_{\star} \la 11.8$ & $\sim700$ & Single S\'{e}rsic & No \\

\citet{Kelkar2015}\tablenotemark{k}& $0.4<z<0.8$ & $10.2\la \log M_{\star} \la 11.7$ & $\sim500$ & Single S\'{e}rsic & No \\

\tableline 
High Redshift, $z\ga1$\\

\citet{Rettura2010}\tablenotemark{l}& $z\sim1.2$ & $10.5 \la \log M_{\star} \la 11.2$\tablenotemark{u} & $44$ & Single S\'{e}rsic & No\\

\citet{Papovich2012}\tablenotemark{m}& $z\sim1.6$ & $10.0 \la \log M_{\star} \la 11.3$ & $\sim60$ & Single S\'{e}rsic & Yes\\

\citet{Raichoor2012}\tablenotemark{n}& $z\sim1.3$ & $9.7 \la \log M_{\star} \la 11.2$\tablenotemark{u} & $76$ & Single S\'{e}rsic & Yes (Opposite)\\

\citet{Bassett2013}\tablenotemark{o}& $z\sim1.6$ & $10.3 \la \log M_{\star} \la 11.3$ & $\sim90$ & Single S\'{e}rsic & Yes\\

\citet{Lani2013}\tablenotemark{p}& $0.5<z<2.0$ & $9.8 \la \log M_{\star} \la 11.6$ & $\sim5000$ & Single S\'{e}rsic & Yes\\

\citet{Strazzullo2013}\tablenotemark{q}& $z\sim2.0$ & $\log M_{\star} \sim 10.7$\tablenotemark{u} & $12$ & Single S\'{e}rsic & Yes\\

\citet{Delaye2014}\tablenotemark{r}& $0.8<z<1.5$ & $10.5 \la \log M_{\star} \la 11.5$ & $\sim700$ & Single S\'{e}rsic & Yes\\

\enddata
\tablecomments{Footnotes are special remarks for each study.}

\tablenotetext{a}{\citet{Maltby2010}: they compared cluster galaxies with field galaxies. This study is based on the galaxies with photometric redshifts. Very few objects are at $\log (M_{\star}/M_{\odot})>11.0$  ($\sim20$ Es). }

\tablenotetext{b}{\citet{Nair2010}: the environments of galaxies are defined by the $N$th nearest objects and a group occupation number. Very few early-type galaxies ($\sim60$) are at $\log (M_{\star}/M_{\odot})>11.3$.}

\tablenotetext{c}{\citet{Cappellari2013}: they compared the Coma Cluster with field galaxies in the local universe. Very few early-type galaxies ($\sim10$) are at $\log (M_{\star}/M_{\odot})>11.3$.}

\tablenotetext{d}{\citet{Huertas2013b}: they compared central galaxies with satellite galaxies. The environments of galaxies are defined by halo mass.}

\tablenotetext{e}{\citet{Poggianti2013}: they compared cluster galaxies with field galaxies. Early-type galaxies in the field are slightly larger than the counterparts in the cluster. Very few early-type galaxies ($\sim50$) are at $\log (M_{\star}/M_{\odot})>11.3$.}

\tablenotetext{f}{\citet{Cebrian2014}: the environments of galaxies are defined by stellar mass density inside a fixed aperture. Additionally, they divided sample into cluster and field galaxies. Early-type galaxies are slightly larger ($\sim3.5\%$) in the low-density regions than in high-density regions.}

\tablenotetext{g}{\citet{Zhao2015}: a sample of the brightest cluster galaxies (BCGs) is used in this study. The environments of galaxies are defined by luminosity density.}

\tablenotetext{h}{This study: the environments of galaxies are defined by the 10th nearest objects with photometric redshifts. Massive early-type galaxies are larger in denser environments. }

\tablenotetext{i}{\citet{Cooper2012}: the environments of galaxies are defined by the $N$th nearest objects. Early-type galaxies in higher-density environments tend to be larger than their counterparts in lower-density environments.}

\tablenotetext{j}{\citet{Huertas2013a}: they compared galaxies in groups with those in the field.}

\tablenotetext{k}{\citet{Kelkar2015}: they compared cluster galaxies with field galaxies. Bad fits ($n>6$  or $n<0.2$ or $R_\mathrm{eff} \ge 5"$) were excluded $(20 - 30\%)$. Very few objects are at $\log (M_{\star}/M_{\odot})>11.4$ ($\sim10$ spectroscopic samples).}

\tablenotetext{l}{\citet{Rettura2010}: they compared cluster galaxies with field galaxies. Sample size is small}

\tablenotetext{m}{\citet{Papovich2012}: they compared cluster galaxies with field galaxies. Quiescent galaxies in the cluster are on average larger compared with those in the field. The number of quiescent galaxies is small $(\la 100)$.}

\tablenotetext{n}{\citet{Raichoor2012}: the environments of galaxies are divided into cluster, group, and field. Early-type galaxies tend to be on average compact in denser environments.}

\tablenotetext{o}{\citet{Bassett2013}: the environments of galaxies are defined by the projected distance from the cluster center. Quiescent galaxies in the clusters are on average larger than field galaxies.}

\tablenotetext{p}{\citet{Lani2013}: the environments of galaxies are defined by mainly a fixed physical aperture and the $N$th nearest objects for comparison. Passive galaxies in denser environments are on average larger compared with the counterparts in less dense environments. Size--density relation tends to be stronger for the galaxies with $\log (M_{\star}/M_{\odot})>11.0$ and weakens from $z=2$ to $z=0.5$. }

\tablenotetext{q}{\citet{Strazzullo2013}: they compared cluster galaxies with field galaxies. Passive early-type galaxies in the clusters are on average larger than field counterparts. The number of passive early-type galaxies used to analyze the mass--size relation is very small $(\sim10)$.}

\tablenotetext{r}{\citet{Delaye2014}: they compared cluster galaxies with field galaxies. Quiescent early-type galaxies in the clusters are on average larger than field counterparts.}

\tablenotetext{s}{The stellar masses assume Chabrier or Kroupa IMFs.}

\tablenotetext{t}{Only for early-type, passive, or quiescent galaxies.}

\tablenotetext{u}{They used the Salpeter IMF \citep{Salpeter1955}. So we subtract 0.3 dex from the stellar mass ranges probed in their studies to convert to the Chabrier IMF.}

\label{tabint}
\end{deluxetable*}

\section{Sample} \label{sec:Data}
 Early-type galaxies used in this study are chosen from the catalog of \citet[][hereafter M14]{Mendel2014}. This catalog is based on the catalog of \citet[][hereafter S11]{Simard2011}. S11 carried out 2D surface brightness profile modeling for 1,123,718 galaxies from Sloan Digital Sky Survey Data Release 7 \citep[SDSS DR7;][]{Abazajian2009}. M14 narrowed down the sample by selecting objects that are in the SDSS main sample \citep{Strauss2002} and spectroscopically classified as galaxies. Using the structural parameters from the S11 catalog, M14 estimated the stellar masses of the selected galaxies by spectral energy distribution (SED) fitting. They estimated not only total stellar mass of each galaxy but also the stellar mass of each galaxy component (i.e.,  bulge and disk) individually. The flexible stellar population synthesis (FSPS) model \citep{Conroy2009}, the initial mass function (IMF) of \citet{Chabrier2003}, and the extinction law of \citet{Calzetti2000} are used for the SED models in M14. Details of the fitting method are described in M14.

 Early-type galaxies are classified based on the Korea Institute for Advanced Study Value-Added Galaxy Catalog \citep{Choi2010}. This catalog is based on the large-scale structure sample of the New York University Value-Added galaxy Catalog \citep{Blanton2005}. In this catalog, \citet{Choi2010} classified galaxies into early-type or late-type using $u - r$ color, $g-i$ color gradient ($\Delta(g-i)$), and inverse concentration index, $c$ in the $i$ band. Galaxies are classified as early-type galaxies according to the following criteria: (1) low value of $c$ $(c<0.43)$, since the light is in general centrally concentrated for early-type galaxies; and (2) red $u-r$ and slightly negative $\Delta(g-i)$ as is the case for most early-type galaxies, or blue $u-r$ and positive $\Delta(g-i)$ to allow the selection of blue early-type galaxies \citep[e.g.,][]{Im2001}.\footnote{More specifically, galaxies are classified as early-type galaxies when they are located above boundaries passing through the points (3.5, -0.15), (2.6, -0.15), and (1.0, 0.3) in the $u - r$ ($x$-axis) versus $g-i$ color gradient ($y$-axis) in Figure 1 of \citet{P&C2005}.}

 They also visually inspected objects to improve their classification. They estimated that the completeness and the reliability of the classification is $\sim90\%$. Details about the morphology classification are given in \citet{P&C2005} and \citet{Choi2010}. Even though other studies used different selection criteria from this work, the bias introduced by the different criteria would be small for very massive galaxies, since the most massive galaxies that we concentrate on here are mostly ellipticals. 
 
 We used early-type galaxies with spectroscopic redshifts in the redshift range $0.1\leq z < 0.15$ from the M14 catalog. Several points to be considered for selecting the redshift range are followings: (1) Angular sizes of the galaxies in higher redshift are small, preventing accurate size measurement. The typical effective radius of the early-type galaxies of $\log (M_{\star}/M_{\odot})\sim11.5$ is $\sim10$ kpc, which corresponds to $\sim3\arcsec$ at $z=0.2$. This angular size is not so much larger than the typical seeing size of $\sim1.\arcsec4$ in the SDSS images. (2) We can avoid or minimize size difference coming from different wavelength windows dependent on the redshift \citep[e.g,][]{Vulcani2014}, when we use a small redshift range. Considering these points, we selected the appropriate redshift range that enables us to secure a substantial amount of massive galaxies and is suitable to investigate massive galaxies $\log (M_{\star}/M_{\odot})\ga11.2$ accordingly.  

 In total, 79,184 early-type galaxies were selected. We downloaded the 79,184 $r$-band-corrected image frames from SDSS DR7 data archive server. Using a flag parameter estimated by SExtractor \citep{Bertin1996}, we sorted out objects severely contaminated by other sources, located on image boundaries, or having bad pixels. All the flagged images were discarded except the objects that are mildly blended with other sources. This procedure leaves 73,168 early-type galaxies.\footnote{From the highest-density environment to the lowest-density environment, 292 (out of 985), 1034 (7962), 2380 (29,427), 2081 (35,621), and 229 (5189) galaxies were discarded. We will show in Appendix \ref{Appendix_B} that the exclusion of the discarded objects does not bias the main result of the paper.} Additionally, we exclude 52 galaxies for which size measurements failed (see Section \ref{sec:Radius}). Consequently, 73,116 early-type galaxies constitute the final sample for this study. To show the characteristics of our sample, we present the stellar mass distribution of our sample of early-type galaxies as a function of environment, their color--mass diagram, bulge-to-total light ratio (B/T) distribution, and axis ratio distribution in Appendix \ref{Appendix_A}.

\begin{figure}
\includegraphics[scale=0.23,angle=00]{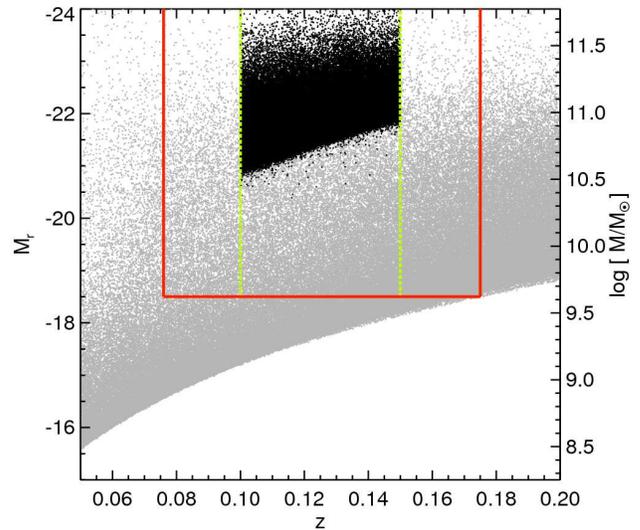}
\centering
	\caption{Early-type galaxies (the black dots) and the galaxies used for the environment measurement (the gray dots) in the redshift-absolute $r$-band magnitude space. The stellar mass is also indicated on the y-axis on the right, assuming the light--mass relation (Equation \ref{eq3}) of early-type galaxies. So, the stellar mass values are only applicable for early-type galaxies. The redshifts of the early-type galaxies are spectroscopic redshifts, while those of the galaxies used for the environment measurement are photometric redshifts. Only 150,000 galaxies used for the environment measurement are shown here for display. The green vertical dashed lines indicate the redshift cut for the early-type galaxies. The red lines represent the boundary for the volume-limited sample of the galaxies used for the environment measurement.
		\label{cutfig}}
\end{figure}

 We used the same $K$-correction\footnote{$K$-corrections were computed using kcorrect software \citep{B&R2007}.} and Galactic extinction\footnote{Galactic extinction values are based on \citet{Schlegel1998}.} values as used in the S11 catalog for the absolute magnitude of the early-type galaxies. Figure \ref{cutfig} shows the absolute $r$-band magnitude (and stellar mass) versus redshift of the early-type galaxy sample (the black dots). We note that our early-type galaxies are volume limited at $\log (M_{\star}/M_{\odot})>10.9$ over $0.1<z<0.15$. Since our main conclusion is based on galaxies with $\log (M_{\star}/M_{\odot})\ga11.2$, the result is not affected by the sample limit. Moreover, due to the very small redshift range, the spatial resolution change as a function of redshift hardly affects our result even if we include galaxies of $\log (M_{\star}/M_{\odot})<10.9$. For galaxies with  $\log (M_{\star}/M_{\odot})\sim10.9$, we examined whether the physical size measurements are biased due to a loss in the image spatial resolution for galaxies at $z=0.15$ in comparison to $z=0.1$. No significant change in the median physical sizes was detected ($\sim 1$ \%), suggesting that the physical size measurements are robust against the image spatial resolution change as a function of redshift.
\\

\begin{figure}
\includegraphics[scale=0.145,angle=00]{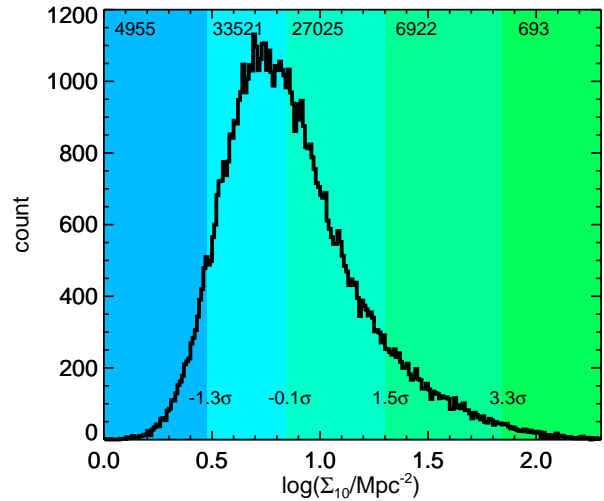}
\centering
	\caption{Surface number density distribution around our early-type galaxies. The bin size is 0.01 dex. We divided these galaxies into five enviroments in this study (colored regions): (1) $0\leq\Sigma_{10}<3$, (2) $3\leq\Sigma_{10}<7$, (3) $7\leq\Sigma_{10}<20$, (4) $20\leq\Sigma_{10}<70$, and (5) $\Sigma_{10}\geq70$, where the unit of surface galaxy density is $\mathrm{Mpc}^{-2}$. The numbers of galaxies in each environment are 4955, 33,521, 27,025, 6922, and 693, respectively. The boundaries correspond to $-1.3\sigma, -0.1\sigma, 1.5\sigma$, and $3.3\sigma$, respectively. 
		\label{denhistfig}}
\end{figure}

\begin{figure*}
\includegraphics[scale=0.1,angle=00]{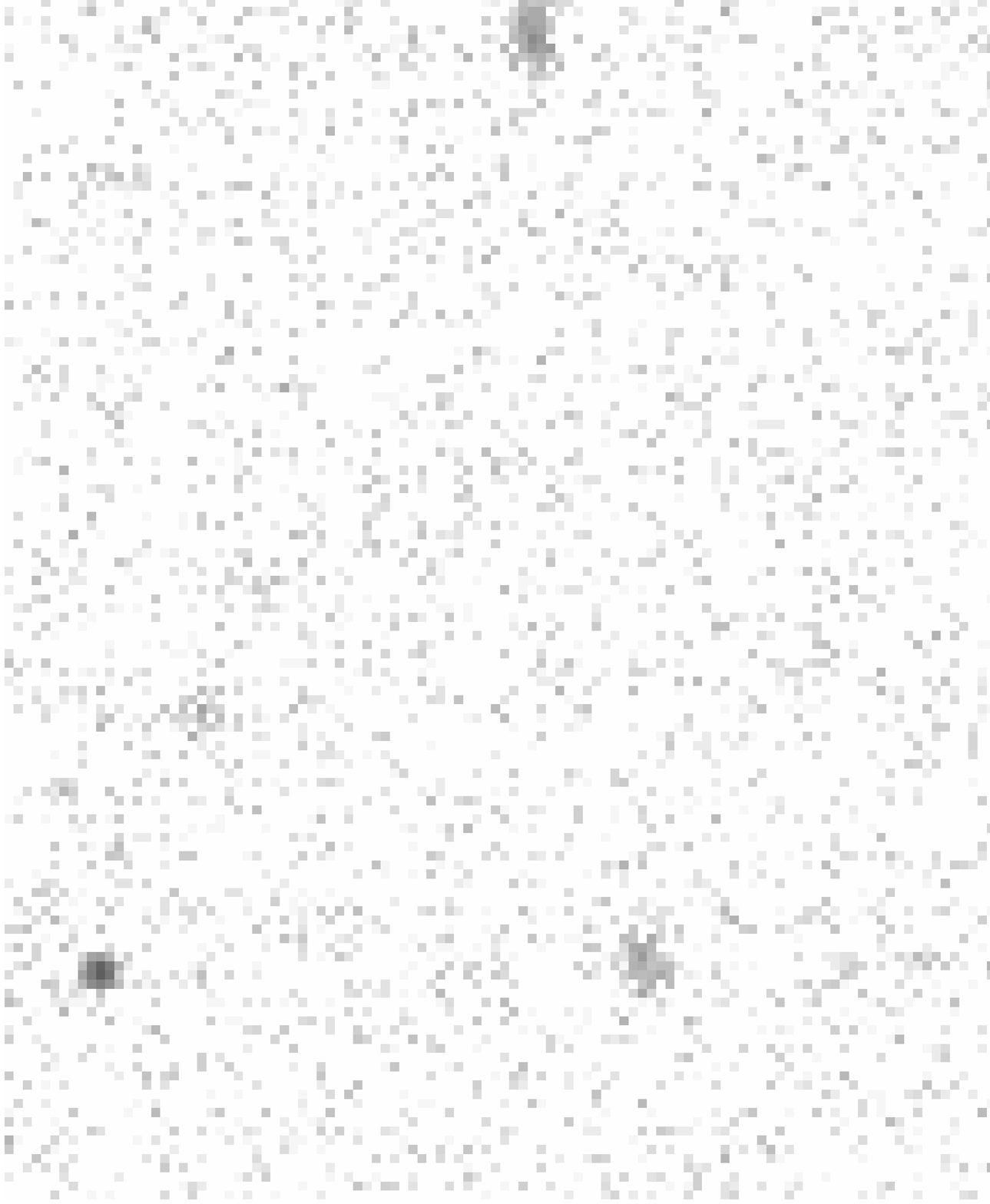}
\centering
	\caption{Galaxies in the highest-density bin. Middle panels show examples of masking processes applied to images in the left panels.  The right panels show examples of the curve of growth for each galaxy. The thick horizontal lines represent the total magnitudes of each galaxy. The vertical dashed lines indicate the effective radii of each galaxy that are seeing corrected by an empirical relation. We showed the radii where total fluxes are measured as the vertical solid lines 
		\label{examfig}}
\end{figure*}

\begin{figure*}
\includegraphics[scale=0.1,angle=00]{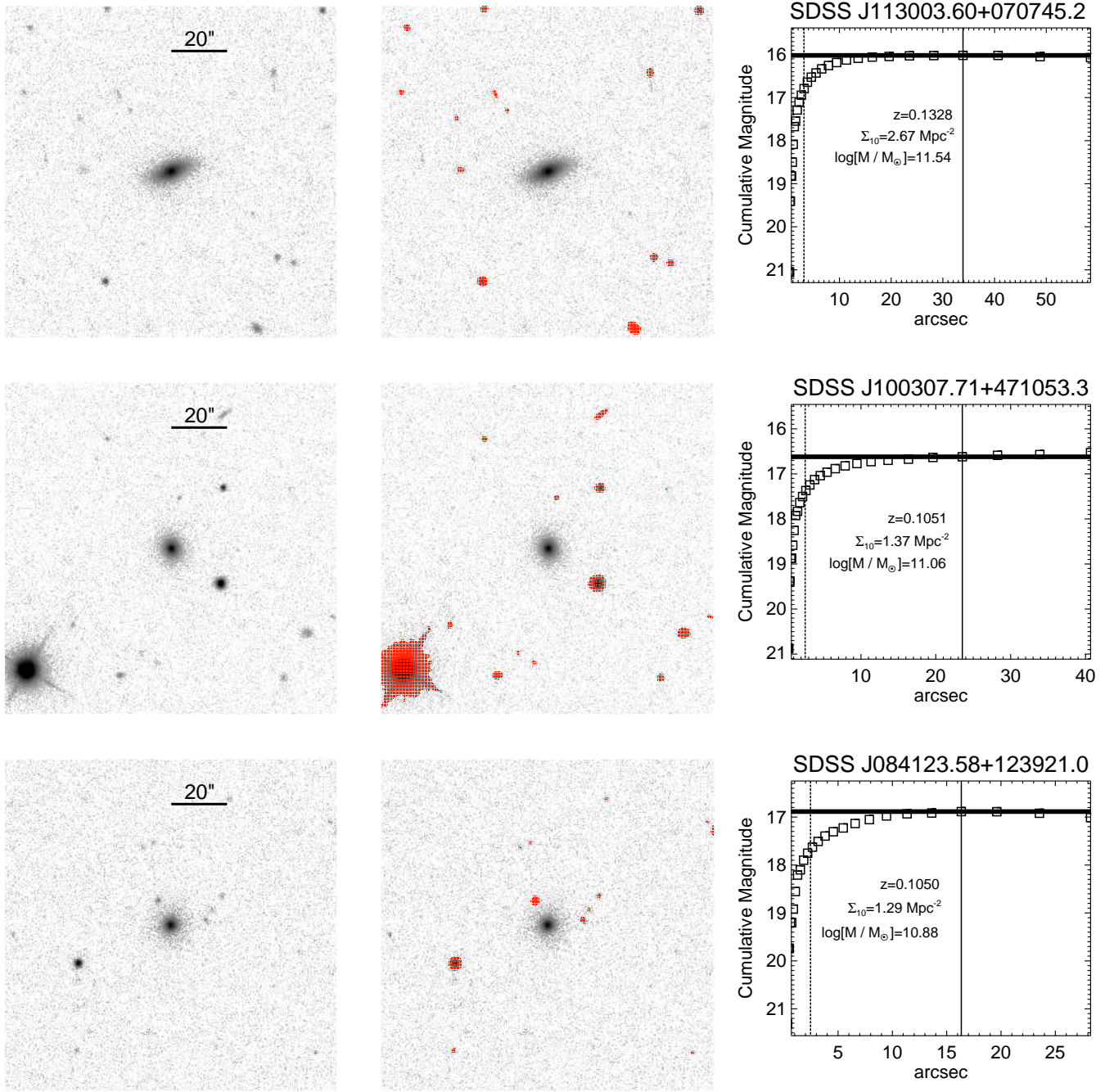}
\centering
	\caption{Same as Figure~\ref{examfig}, but for the galaxies in the lowest-density bin.
		\label{exam2fig}}
\end{figure*}

\section{Analysis} \label{sec:Analysis}
\subsection{Environment Measurements}  \label{sec:Environment}
 To define the environment around each early-type galaxy, we used a volume-limited sample of all photometric objects classified as galaxies and their photometric redshifts\footnote{The photometric redshifts are measured by a hybrid method that is based on both template-fitting approach and empirical calibration using galaxies with colors and spectroscopic redshifts \citep{Abazajian2009}.} in the SDSS DR7 with $M_r < -18.5$ mag.  

 Figure \ref{cutfig} also shows galaxies used for the environment measurement. The green vertical dashed lines indicate the redshift cut for the early-type galaxies. The red lines represent the boundary for the volume-limited sample of galaxies. The vertical red lines are set at 6500km s$^{-1}$ away from the redshift cuts for the early-type galaxy sample ($z=0.1$ or $0.15$) to account for the photometric redshift error (see below). Note that $m_r=21.2$ mag, which is $M_r=-18.5$ mag at $z=0.175$ (the highest redshift of the galaxies used for the environment measurement), is the completeness limit of $\sim90\%$ for detection and $\sim95\%$ for source classification \citep{Wang2013}.

 We defined the environment as the surface number density of the galaxies within the 10th nearest neighbor, using the galaxies whose photometric redshifts fall between velocity cuts ($\pm6500$km s$^{-1}$) centered on the spectroscopic redshifts of early-type galaxies, where $dv=\pm6500$km s$^{-1}$ corresponds to typical rms values of the difference between photometric redshifts and spectroscopic redshifts in the explored redshift range as measured by us. Then the galaxy surface number density is calculated by
\begin{equation}
	\Sigma_{n}=\frac{n}{\pi {r_{n}}^2},
\label{eq1}
\end{equation}
 where $r_{n}$ is projected distance to the $n$th nearest neighbor and $n=10$ in this study. 

 Figure \ref{denhistfig} shows the surface number density distribution around our early-type galaxies. We divided the surface number densities into five bins to define different environments: (1) $0\leq\Sigma_{10}<3$, (2)  $3\leq\Sigma_{10}<7$, (3) $7\leq\Sigma_{10}<20$, (4) $20\leq\Sigma_{10}<70$, and (5) $\Sigma_{10}\geq70$. The numbers of galaxies in each environment are 4955, 33,521, 27,025, 6922, and 693, respectively. The boundaries correspond to $-1.3\sigma, -0.1\sigma, 1.5\sigma$, and $3.3\sigma$, respectively, where $\sigma$ is a standard deviation of the surface number densities that was derived after 3$\sigma$ clipping with maximum iteration of 5. We selected each boundary to secure enough galaxies in the highest- and the lowest-density bins for robust statistics.
 
 We also tried the local densities that are defined differently: the surface number densities within the fifth, seventh nearest-neighbor galaxies, the luminosity density within fixed aperture of 1 Mpc, and the luminosity density within the 10th nearest objects (all with the same velocity cut). Note that the luminosity densities are tried as a proxy for the mass density, which is suggested as a good measure of the local density \citep[e.g.,][]{Wolf2009}. As we shall show later, there is no meaningful change in the main conclusion of the paper due to adopting a different local density notation; thus, we will present only the results that are based on $\Sigma_{10}$.
\\

\begin{figure}
\includegraphics[scale=0.200,angle=00]{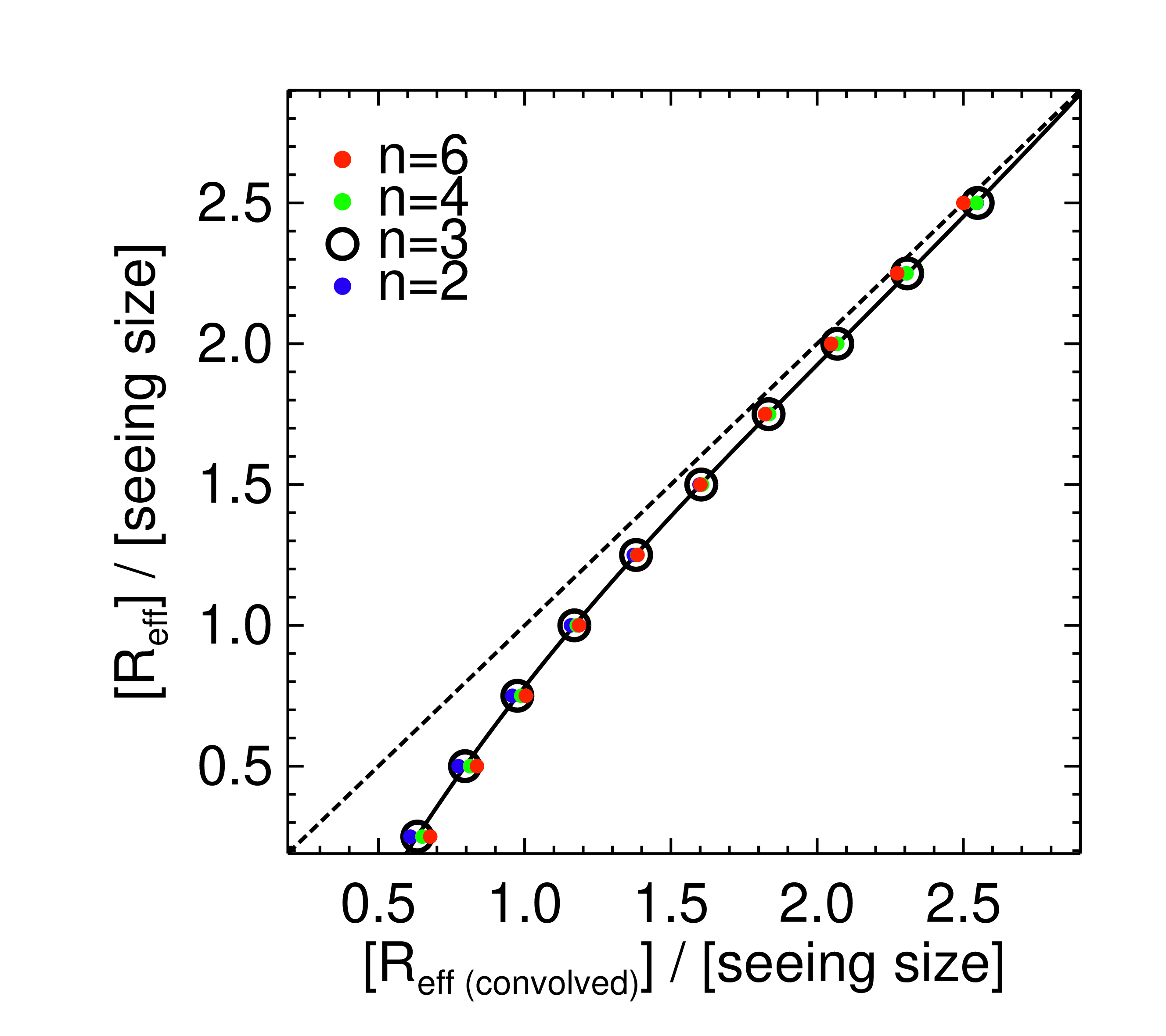}
\centering
	\caption{Empirical relation for seeing correction in size measurement. We used S\'{e}rsic models with S\'{e}rsic indices of $n=2,3,4$, and $6$. We fit the points for $n=3$ with a fourth-order polynomial function to construct the relation. The dashed line indicates $R_\mathrm{eff}=R_\mathrm{eff(convolved)}$.
		\label{seeingfig}}
\end{figure}

\begin{figure}
\includegraphics[scale=0.27,angle=00]{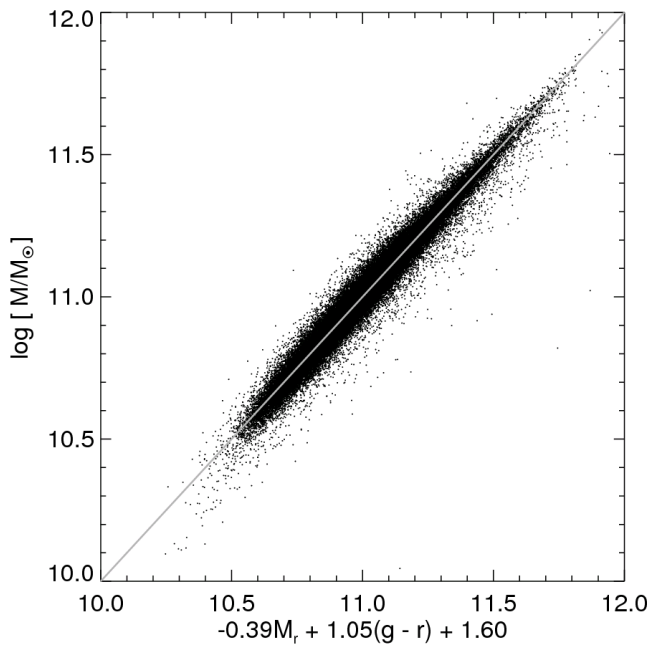}
\centering
	\caption{Comparison between the values derived by the relation and the stellar masses. $M_r$ is the absolute magnitude of $r$ band, and $g-r$ is the color of the galaxy at $z=0$. The gray line denotes a one-to-one relation.
		\label{mlfig}}
\end{figure}

\begin{figure*}
\includegraphics[scale=0.12,angle=00]{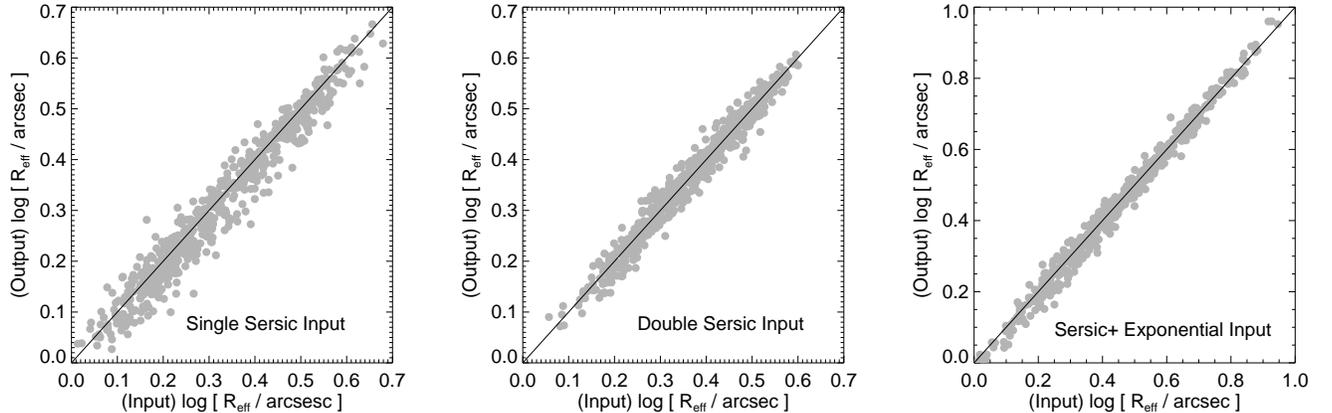}
\centering
	\caption{Comparison between input sizes of simulated galaxies and their sizes measured by the nonparametric method. The left panel is for the single S\'{e}rsic model, the middle panel is for the double S\'{e}rsic model, and the right panel is for the S\'{e}rsic + exponential model. The lines denote one-to-one relations.
		\label{compfig}}
\end{figure*}

\begin{figure*}
\includegraphics[scale=0.12,angle=00]{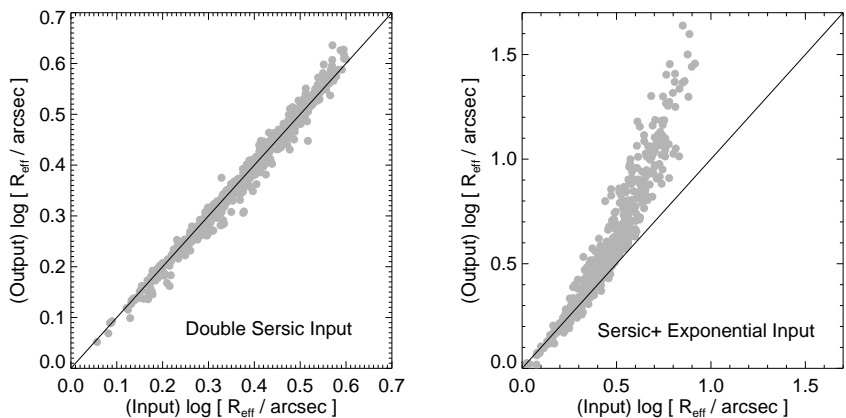}
\centering
	\caption{Comparison between input sizes of simulated galaxies and their sizes measured by the single S\'{e}rsic fit. The left panel is for the double S\'{e}rsic model, and the right panel is for the S\'{e}rsic + exponential model. The lines denote one-to-one relations.
		\label{compmodelfig}}
\end{figure*}

\begin{figure*}
\includegraphics[scale=0.18,angle=00]{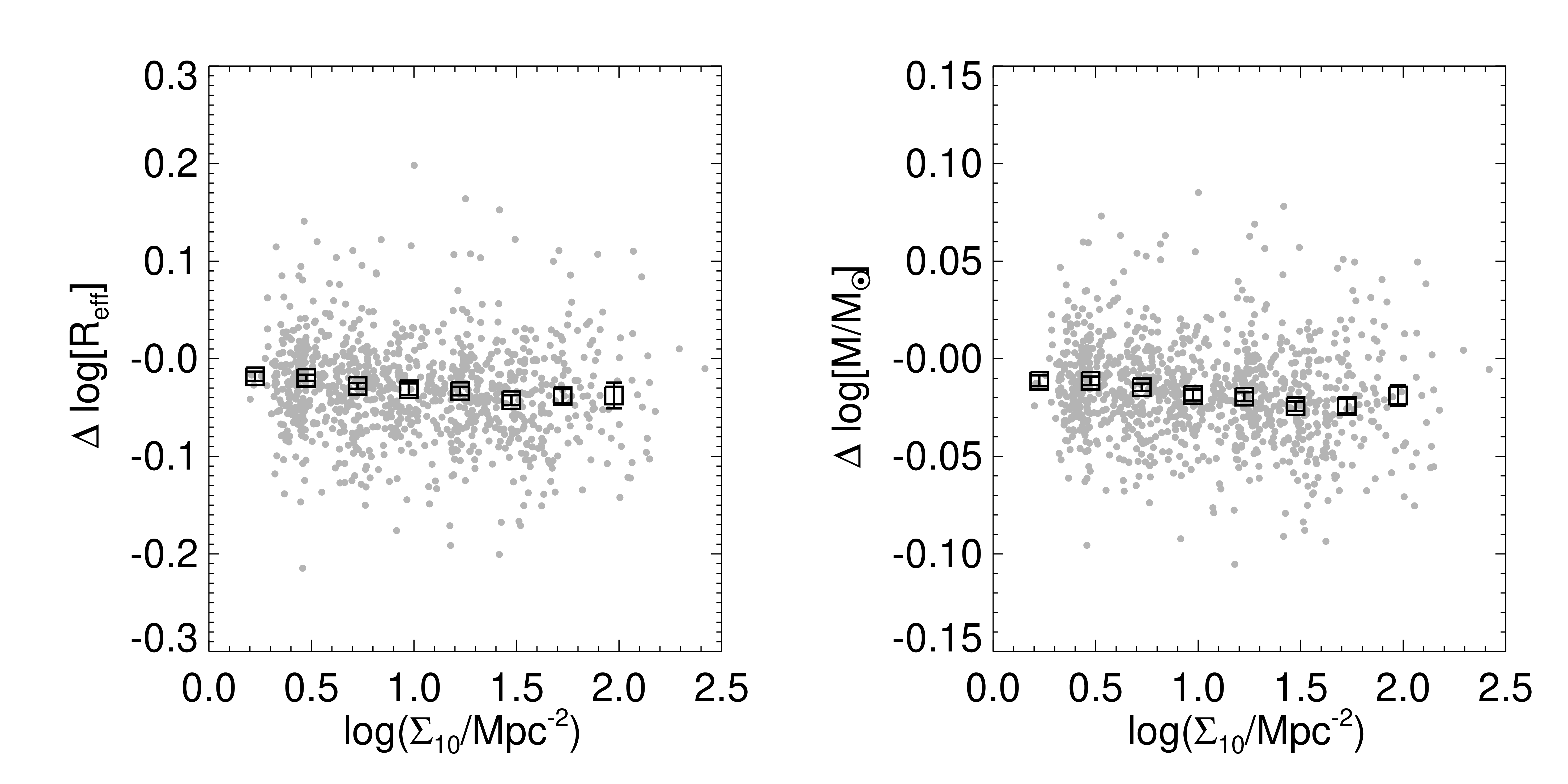}
\centering
	\caption{We tested how neighboring sources affect the size and total flux measurement. This figure indicates differences of the measured size (left) and stellar mass (right) as a function of environment, in which the differences are between two simulated cases (with/without all sources in the environment). Negative values mean underestimated ones for the case with all sources in the environment. The black squares are median values in each environment bin. The error bar of each square is $1\sigma$ of the median values from 200 bootstrap resampling.
		\label{envfig}}
\end{figure*}

\subsection{Size and Total Flux Measurements}  \label{sec:Radius}
 We used a nonparametric method to measure sizes of galaxies, by constructing curve of growth of each galaxy and finding a radius where one-half of the total flux is contained.

 To construct the curve of growth and derive sizes, we ran IRAF\footnote{IRAF is distributed by the National Optical Astronomy Observatory, which is operated by the Association of Universities for Research in Astronomy (AURA) under a cooperative agreement with the National Science Foundation.} ELLIPSE on $r$-band images (without cutout) from SDSS DR7 (see Section \ref{sec:Data}). Before running ELLIPSE, we masked all sources except the galaxy, whose size we want to measure. Masking files readable by the ELLIPSE were automatically generated based on SExtractor segmentation files. We masked all sources above a detection threshold of $1.5\sigma$ relative to the background and with minimum pixels of 2 above that threshold. The middle panels of Figures \ref{examfig} and \ref{exam2fig} show examples of masked regions applied to images in the left panels. Figure \ref{examfig} is for galaxies in the highest-density bin, while Figure \ref{exam2fig} is for galaxies in the lowest-density bin. 
 
We then ran ELLIPSE for the galaxies. We fit elliptical isophotes in such a way that a semimajor axis (SMA) increases 1.2 times the previous SMA of an elliptical isophote. The isophotes are fit up to the maximum semimajor axis length, which is set to be 24 times the detected radius of the galaxy in the SExtractor segmentation image. For the initial input values of the fitting parameters such as ellipse center coordinate, ellipticity, and position angle, we used the values of the de Vaucouleurs fit from the SDSS DR7 photometry object table. Then all these fitting parameters were set free during the fitting process.

 The background value was determined by an average pixel value in the circular annulus whose inner radius is $\sim12$ times and outer radius is $\sim24$ times the detected size of the galaxy in the SExtractor segmentation image, while all sources were masked.

 After subtracting the background value by this method, we derived the cumulative flux as a function of circular radius\footnote{We used a cumulative flux within a circular radius to derive a circularized effective radius.} (curve of growth) from the galaxy center. Then we derived the slope of the curve of growth and found the minimum of the absolute value of the slope. The radius where the absolute value of the slope is minimum was set to be the radius of the aperture within which the total flux of the galaxy was measured. 
Neighboring bright sources could affect the curve of growth severely in such a way that the galaxy total flux would be overestimated because the outer, unmasked regions of the neighbor sources could be counted in the curve of growth. To avoid such a bias, we set an upper limit on the aperture size as the distance to the nearest bright source (brighter than the magnitude of the sample galaxy $+2$) and determined the minimum of the absolute value of the slope in that limit. Furthermore, we set the lower limit of the aperture size to be 3 times the detected size of the galaxy in the SExtractor segmentation image. This is to prevent the situation in which substantial flux from the object falls outside of the aperture when the bright sources are quite close to the object. 

 Using the total flux and the curve of growth, circularized effective radius\footnote{We omit ``circularlized" from now on.} ($R_\mathrm{eff}$) was easily derived by its definition: the radius within which half of the total flux is contained. We used the effective radius as the size of a galaxy. The right panels of Figures \ref{examfig} and \ref{exam2fig} show examples of the curve of growth.\\

 The central part of the galaxy surface brightness profile is smeared due to seeing, and this can affect the size measurement. We derived a simple empirical relation to correct for the smearing effect in the size measurement. For this, we made a set of single S\'{e}rsic galaxy models with or without seeing convolution. The models were generated by the GALFIT software \citep{Peng2010}, where the seeing smearing is approximated with a gaussian point spread function. We measured the sizes of the simulated galaxies. We then derived a relation between the effective radii of the seeing-convolved galaxies ($R_\mathrm{eff(convolved)}$) and those without seeing convolution ($R_\mathrm{eff}$) as shown in Figure \ref{seeingfig}. Fitting the points for $n=3$ with a fourth-order polynomial function, we obtained the following relation:
\begin{equation}
	y = -0.960 + 2.305x - 0.723x^2 + 0.175x^3 - 0.014x^4,
\label{eq2}
\end{equation}
 where $x$ is $R_\mathrm{eff(convolved)}$, and $y$ is $R_\mathrm{eff}$. 

 Early-type, bulge-dominated galaxies generally have S\'{e}rsic indices in the range of $2\lesssim n \lesssim6$ \citep{Fisher2008, Fisher2010}. We constructed the relations for the S\'{e}rsic models with S\'{e}rsic indices of $n=2,3,4$, and $6$. As shown in Figure \ref{seeingfig}, they do not show significant differences from each other. Therefore, we adopted the relation for $n=3$ in this study. Use of the other S\'{e}rsic indices does not change our results. 

If $R_\mathrm{eff(convolved)} > \sim3$ times the seeing size, the relation converges to $R_\mathrm{eff} = R_\mathrm{eff(convolved)}$ at $99.98\%$ accuracy, since effective radii of large galaxies are hardly affected by the seeing effect. To correct for the seeing effect in the size measurement, we applied this relation to the galaxies whose measured effective radii are smaller than 3 times their seeing sizes.\footnote{We used the seeing sizes listed in the headers of the images.} All sizes presented in this paper are the ones corrected for the seeing effect. \\

 In this study we used our own stellar masses converted from the total magnitudes measured by our method, because sizes of galaxies are coupled with galaxy total magnitudes.
To convert the total flux into stellar mass, we constructed a tight relation by using the properties of early-type galaxies from S11 and M14 catalogs.
The relation is
\begin{equation}
	\log (M_{\star}/M_{\odot}) = -0.39M_r + 1.05 (g-r) +1.60,
\label{eq3}
\end{equation}
where $M_r$ is the absolute magnitude of $r$ band and $g-r$ is the color of the galaxy at $z=0$. The $g-r$ values were taken from the S11 catalog. Figure \ref{mlfig} shows the comparison between the values derived by the relation and the stellar masses, in which the average offset between two values is 0.00 dex and the standard deviation of the offsets is 0.04 dex.\\

We tested our method to measure galaxy sizes in the following ways. First, we used simulated galaxies to test whether measured sizes coincide with the input sizes of the simulated galaxies. Second, we tested how neighboring sources affect the size and total flux measurement and what biases they could cause in the mass--size relation. Here, input sizes in the simulation are the circularized effective radii as directly measured from the model images before the PSF convolution using the same nonparametric method.

For the first test, we created three sets of 600 simulated galaxies by the GALFIT software. The simulated galaxies in the first set were generated by using a single S\'{e}rsic model. We set the galaxies to have random properties in similar ranges of values to real galaxies: (1) $15.5\le m_r <17.5$, (2) $0.1\le z<0.15$, (3) $2\le n<8$, and (4) $0.4\le b/a<1$. We used the simulated galaxies to follow the luminosity--size relation of \citet[Figure 6]{Shen2003} for early-type galaxies. 

The second set is modeled by a double S\'{e}rsic model. In the second set, the simulated galaxies were generated by the randomly combined two simulated galaxies of the first set. 

The third set is modeled by a S\'{e}rsic + exponential model. We used the same simulated galaxies in the first set for S\'{e}rsic components except that they have magnitudes in the range of $16.2\le m_r <18.2$. For the exponential components, we also used the same simulated galaxies in the first set except that they have a fixed S\'{e}rsic index of 1 and magnitudes in the range of $16.2\le m_r <18.2$. An additional exception was that the exponential components have $1.5$ -- $4$ times larger effective radii than those of single S\'{e}rsic models, which is a typical range of the values of the two-component decomposed models in the S11 catalog.

We set all the images of the simulated galaxies to have various seeing and noise values that are typical of SDSS images. 

We measured the sizes of the simulated galaxies in the three sets by the nonparametric method. The comparison between input sizes of the simulated galaxies of each set and their measured sizes are shown in Figure \ref{compfig}. The figure indicates that our nonparametric method well recovers the sizes of the simulated galaxies regardless of the input models. Moreover, smearing effect due to seeing is well corrected for small galaxies. The mean difference between input and the measured sizes is smaller than $\sim0.01$ dex for the three sets, and the mean absolute deviation between them is smaller than $\sim0.04$ dex for the three sets.

For comparison, we measured the sizes of the simulated galaxies by fitting the single S\'{e}rsic model which has been used by many other studies. The comparison between input sizes of the simulated galaxies of the two sets (double S\'{e}rsic and  S\'{e}rsic + exponential model) and their measured sizes are shown in Figure \ref{compmodelfig}. For the case of the double S\'{e}rsic model, the sizes are well recovered. On the other hand, the sizes are overestimated for the S\'{e}rsic + exponential model. It is more severe for the large galaxies. We find that the overestimates occur when the model tries to recover the outer exponential profile by increasing the S\'{e}rsic indices, sometimes as large as $n=8$. It has already been noted by previous studies that the use of the single S\'{e}rsic model yields overestimated sizes for large galaxies \citep{Mosleh2013,Bernardi2014} when the input surface brightness profile of the galaxies is the S\'{e}rsic + exponential model, which is probably the most realistic surface brightness profile model \citep{Bernardi2013,Meert2013,Bernardi2014}.\\

The second test is to understand how neighboring sources in the images affect the size and the total flux measurements. From our sample SDSS images, we selected 1000 samples to have full ranges of mass and environment. Then we extracted information about all sources in each image by using SExtractor. After that, this information was implemented to the simulated images by GALFIT software with seeing convolution and addition of typical noise of the SDSS images. The stars were modeled by a gaussian model, and the galaxies were modeled by a single S\'{e}rsic model with random S\'{e}rsic indices in a range of $2\le n<8$. The class star parameter of SExtractor is used to divide the stars and galaxies with a separation value of 0.5. Other specific galaxy parameters such as position angles, axis ratios, and sizes (half-light radii) were from the information extracted by the SExtractor. By this process, we made 1000 copies of real images. The advantages of this test using copies of the real images are as follows: (1) we know the input sizes of the target galaxies; (2) spatial and luminosity distributions of the neighboring sources are the same as the real cases. 

Then we measured the sizes of the 1000 simulated galaxies, which are the copies of the sample of the early-type galaxies, by the nonparametric method as described above. For comparison, we generated the same images, but without all sources in the environment. Then we measured their sizes in the same manner. Figure \ref{envfig} shows differences of the measured size and stellar mass as a function of environment, in which the differences are between two simulated cases (with/without all sources in the environment). Negative values mean underestimated ones for the case with all sources in the environment. The sources in the nearby targets cause the size and the mass of the galaxies to be slightly underestimated, since the background value is slightly overestimated by neighbor sources even after they were all masked. Thus, galaxies in high-density environments are more influenced by such effects. 

 The size difference between the galaxies in the lowest-density environments and those in the highest-density environments is $\sim0.02$ dex, and the mass difference between them is $\sim0.01$ dex. This minor bias across the environment cannot alter the results shown in Section \ref{sec:ms}. Even if the bias is corrected in the mass--size relation, the environmental dependence of the mass--size relation increases. 
\\

\begin{figure}
\includegraphics[scale=0.74,angle=00]{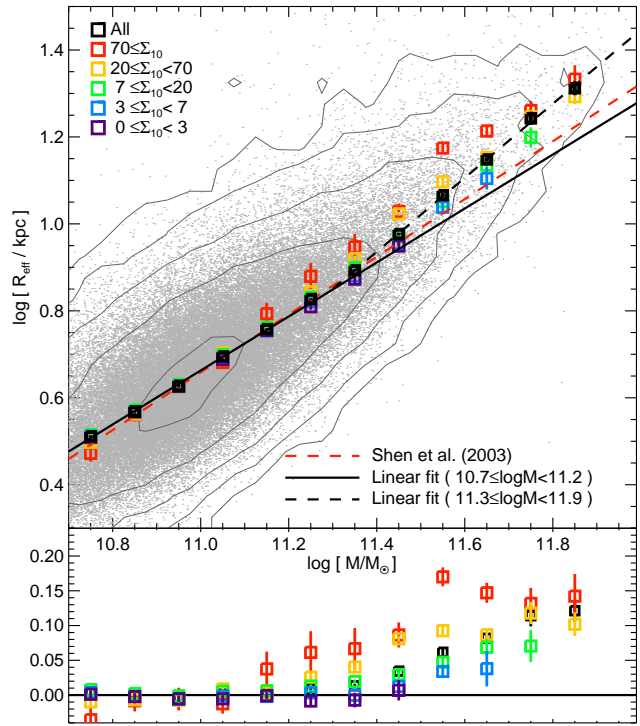}
\centering
	\caption{Mass--size relation of the galaxies with $\log (M_{\star}/M_{\odot})>10.7$. The gray dots in the background denote each galaxy. Contour lines indicate the same density of the gray points and are plotted in a logarithmic scale. The black squares are median effective radii of all galaxies in each mass bin. The size of the mass bin is 0.1 dex. The colored squares are median effective radii of the galaxies under different environments as described in Section \ref{sec:Environment}. The error bar of each square point is $1\sigma$ of the median effective radii from 200 bootstrap resampling. The solid black line is the best-fit mass--size relation for galaxies at $10.7\leq \log (M_{\star}/M_{\odot})<11.2$ and it is extrapolated to $\log (M_{\star}/M_{\odot})>11.2$ for demonstration purposes. Differences between all squares and the linear fit line are shown in the bottom subpanel. The dashed black line is the best-fit mass--size relation for galaxies at $11.3\leq \log (M_{\star}/M_{\odot})<11.9$. For comparison, the linear mass--size relation from \citet[Figure 6]{Shen2003} is shown as a red dashed line.
		\label{ms_1fig}}
\end{figure}

\begin{figure*}
\includegraphics[scale=1.5,angle=00]{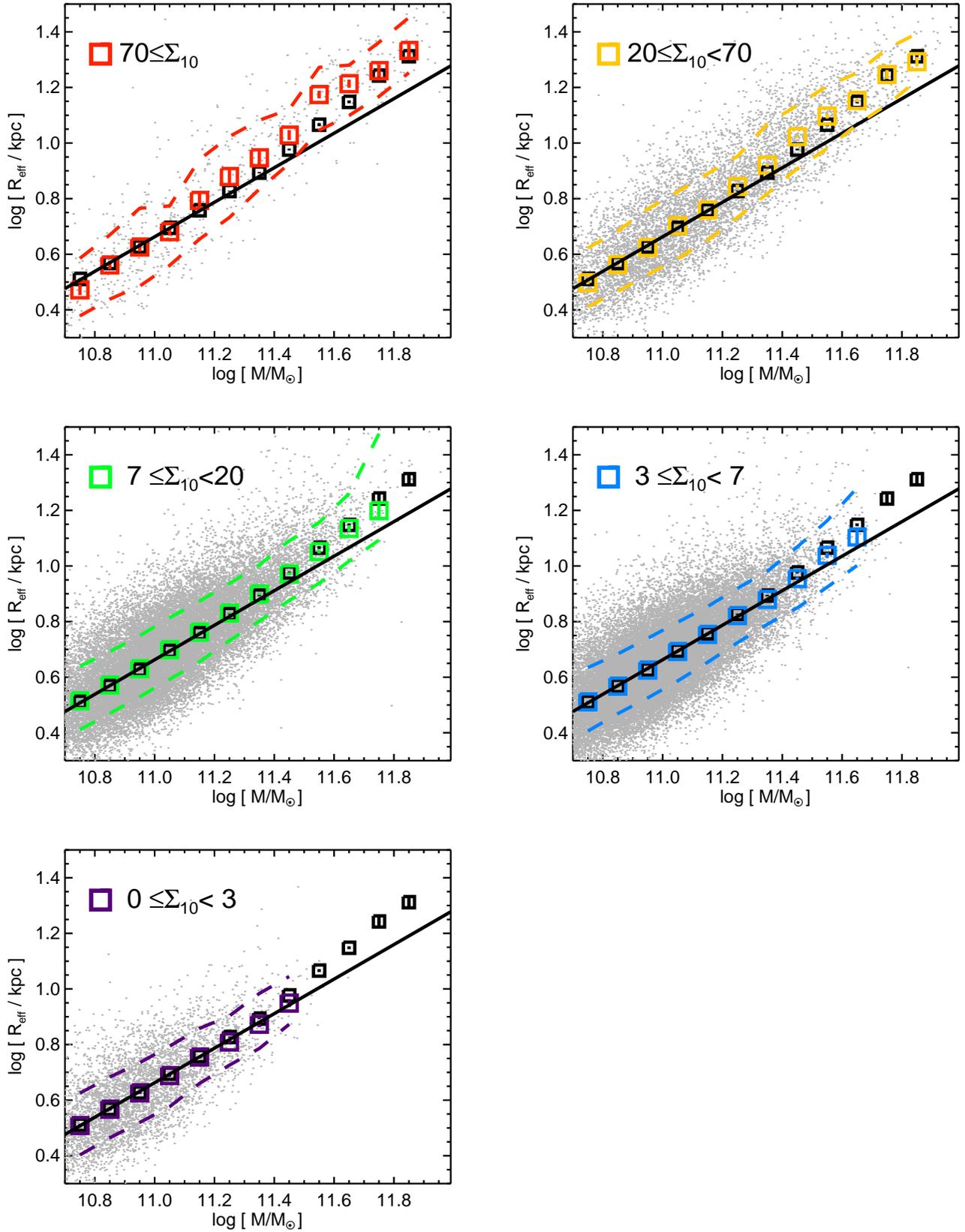}
\centering
	\caption{Mass--size relation of the galaxies with $\log (M_{\star}/M_{\odot})>10.7$. Each environment is separately shown in each panel with the same symbol as in Figure \ref{ms_1fig}. The dashed lines represent the 16th and 84th percentiles ($1\sigma$) of effective radii. 
		\label{ms_2fig}}
\end{figure*}

\begin{figure}
\includegraphics[scale=0.74,angle=00]{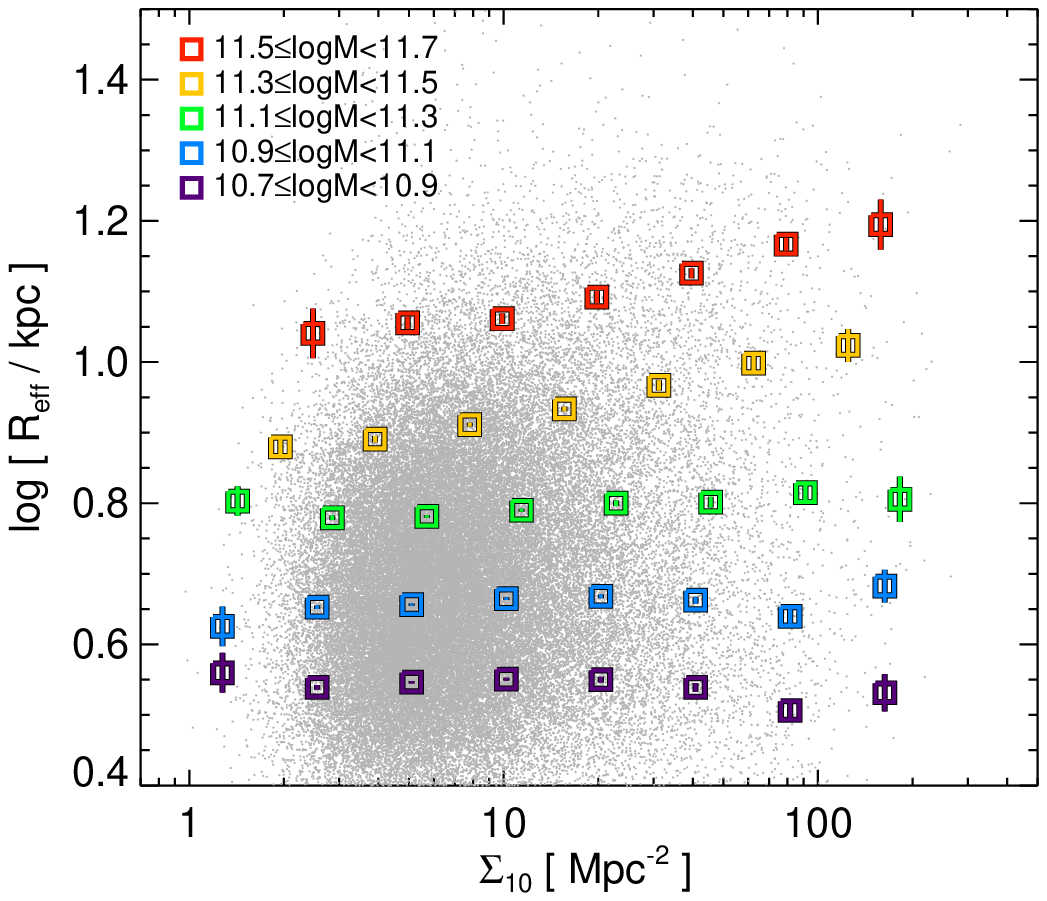}
\centering
	\caption{Environmental dependence of the effective radius in the size--density plane. The gray dots in the background denote whole galaxies. The colored squares are median effective radii of the galaxies under different mass bins. The error bar of each square point is $1\sigma$ of the median effective radii of 200 bootstrap resampling. 
		\label{ms_3fig}}
\end{figure}

\section{Result: Environmental Dependence of Mass--Size Relation} \label{sec:ms}
 Figure \ref{ms_1fig} shows the mass--size relation of early-type galaxies with $\log (M_{\star}/M_{\odot})>10.7$. The gray dots in the background denote galaxies in our sample. The black squares are median effective radii of all galaxies in each mass bin. The colored squares are median effective radii of galaxies in different environments as described in Section \ref{sec:Environment}. In each square point, more than 18 galaxies are included to calculate a median value for statistical robustness. The error bar of each square point is $1\sigma$ of the median effective radii from 200 bootstrap resampling. The solid black line is the best-fit mass--size relation for galaxies at $10.7\leq \log (M_{\star}/M_{\odot})<11.2$ and it is extrapolated to $\log (M_{\star}/M_{\odot})>11.2$ for demonstration purposes. The equation of this solid black line is
\begin{equation}
	\log(R_\mathrm{eff}/\mathrm{kpc}) = 0.621 \log (M_{\star}/M_{\odot}) - 6.169. 
\label{eq4}
\end{equation}
 Differences between all squares and the linear fit line are shown in the bottom subpanel. The dashed black line is the best-fit mass--size relation for galaxies at $11.3\leq \log (M_{\star}/M_{\odot})<11.9$. The equation of this dashed black line is
\begin{equation}
	\log(R_\mathrm{eff}/\mathrm{kpc}) = 0.851 \log (M_{\star}/M_{\odot}) - 8.770. 
\label{eq5}
\end{equation}
Figure \ref{ms_2fig} shows galaxies in different environments separately and their mass--size relations. The mass--size relation is summarized in Table \ref{tab1}. 

 The red dashed line in Figure \ref{ms_1fig} is the mass--size relation of the early-type galaxies from \citet[Figure 6]{Shen2003}, which is applicable at $\log (M_{\star}/M_{\odot})<11.6$. Here, we converted their luminosity--size relation based on $r$ band and the S\'{e}rsic fit into the mass--size relation by converting $M_r$ to mass using Equation \ref{eq3} in Section \ref{sec:Radius}. For $g-r$, the median color of the early-type galaxies from the S11 catalog was used. Their relation agrees well with our mass--size relation in the mass range of $10.7< \log (M_{\star}/M_{\odot})<11.4$. 

 Some studies revealed that the most massive part ($M_{\star}>2\times10^{11}\,M_{\odot}$) of the mass--size relation is curved upward \citep{Desroches2007,Hyde2009,Bernardi2011b}. Our result also shows the curved mass--size relation. A newly discovered fact here is that this trend is more severe for the galaxies in denser environments.

At $10.7\leq \log (M_{\star}/M_{\odot})<11.2$, the environmental dependence of the mass--size relation is negligible. However, the environmental dependence becomes notable and stronger as the mass increases at $\log (M_{\star}/M_{\odot})>11.2$, in such a way that galaxies become larger at higher-density environments. Galaxies in the densest environment are the largest by $\sim0.1$ dex ($25\%$) in comparison to galaxies in the lowest-density environment. Figure \ref{ms_3fig} shows this environmental dependence of the sizes in a size--density plane. Here, the environmental dependence is examined out to $\Sigma_{10}\sim100$, and we see that the most massive galaxies are $\sim0.15$ dex ($40\%$) larger in size at $\Sigma_{10}\sim100$ than at $\Sigma_{10}\sim10$.
 
 To show that the environmental dependence of the mass--size relation is robust, we constructed the mass--size relation using differently defined sizes and local densities and a sample that includes objects with bright neighbor galaxies. For example, we used half-light semimajor axis instead of the circularized effective radius since the major-axis defined size is possibly a more fundamental quantity \citep[e.g.,][]{Trippe2016}, and the local luminosity densities as a proxy for mass densities since several works suggest that the mass densities are better environment measures. We also took stellar masses and effective radii from S11 and M14 and included in the analysis the galaxies we excluded in the nonparametric method due to a blending problem with bright neighbors. We also constructed a sample where stellar mass distribution in each bin is matched with each other, since a slight difference in the mass distribution in each mass bin might introduce a bias. The results of these analyses are presented in Appendix \ref{Appendix_B}, which shows the environmental dependence in the mass--size relation at a similar level to Figures \ref{ms_1fig} - \ref{ms_3fig}. Therefore, we conclude that the environmental dependence of the mass--size relation is genuine, not an artifact due to a special way that the sample is constructed or analyzed.
\\

\begin{deluxetable*}{ccccccc}
\tablecaption{Median Effective Radii of Galaxies [$\log(R_\mathrm{eff}/\mathrm{kpc})$]} 
\tabletypesize{\scriptsize}
\tablehead{
\colhead{Stellar Mass} & \colhead{All Galaxies} & \colhead{$0\leq\Sigma_{10}<3$} & \colhead{$3\leq\Sigma_{10}<7$} & \colhead{$7\leq\Sigma_{10}<20$} & \colhead{$20\leq\Sigma_{10}<70$} & \colhead{$\Sigma_{10}\geq70$}
}
\startdata
$10.7\leq \log (M_{\star}/M_{\odot})<10.8$&$0.511\pm0.001$&$0.509\pm0.006$&$0.511\pm0.002$&$0.515\pm0.003$&$0.498\pm0.006$&$0.472\pm0.019$ \\
$10.8\leq \log (M_{\star}/M_{\odot})<10.9$&$0.568\pm0.001$&$0.567\pm0.005$&$0.568\pm0.002$&$0.572\pm0.003$&$0.563\pm0.005$&$0.561\pm0.015$ \\
$10.9\leq \log (M_{\star}/M_{\odot})<11.0$&$0.627\pm0.001$&$0.626\pm0.004$&$0.626\pm0.002$&$0.631\pm0.001$&$0.624\pm0.004$&$0.626\pm0.016$ \\
$11.0\leq \log (M_{\star}/M_{\odot})<11.1$&$0.695\pm0.001$&$0.688\pm0.005$&$0.692\pm0.001$&$0.699\pm0.002$&$0.703\pm0.005$&$0.681\pm0.014$ \\
$11.1\leq \log (M_{\star}/M_{\odot})<11.2$&$0.758\pm0.001$&$0.755\pm0.005$&$0.754\pm0.002$&$0.762\pm0.002$&$0.759\pm0.007$&$0.794\pm0.025$ \\
$11.2\leq \log (M_{\star}/M_{\odot})<11.3$&$0.827\pm0.002$&$0.809\pm0.007$&$0.821\pm0.003$&$0.831\pm0.002$&$0.844\pm0.007$&$0.879\pm0.031$ \\
$11.3\leq \log (M_{\star}/M_{\odot})<11.4$&$0.894\pm0.002$&$0.873\pm0.011$&$0.880\pm0.003$&$0.900\pm0.003$&$0.920\pm0.006$&$0.947\pm0.030$ \\
$11.4\leq \log (M_{\star}/M_{\odot})<11.5$&$0.977\pm0.003$&$0.949\pm0.015$&$0.955\pm0.005$&$0.972\pm0.006$&$1.023\pm0.004$&$1.029\pm0.018$ \\
$11.5\leq \log (M_{\star}/M_{\odot})<11.6$&$1.066\pm0.005$&\nodata&$1.038\pm0.009$&$1.052\pm0.006$&$1.097\pm0.009$&$1.174\pm0.014$ \\
$11.6\leq \log (M_{\star}/M_{\odot})<11.7$&$1.148\pm0.005$&\nodata&$1.104\pm0.025$&$1.135\pm0.012$&$1.153\pm0.006$&$1.214\pm0.014$ \\
$11.7\leq \log (M_{\star}/M_{\odot})<11.8$&$1.243\pm0.015$&\nodata&\nodata&$1.199\pm0.023$&$1.246\pm0.016$&$1.260\pm0.022$ \\
$11.8\leq \log (M_{\star}/M_{\odot})<11.9$&$1.312\pm0.016$&\nodata&\nodata&\nodata&$1.293\pm0.017$&$1.333\pm0.032$ \\
\enddata
\tablecomments{The unit of surface galaxy density $\Sigma_{10}$ is $\mathrm{Mpc}^{-2}$. The errors are $1\sigma$ of the median effective radii from 200 bootstrap resampling.}
\label{tab1}
\end{deluxetable*}

\section{Discussion} \label{sec:Discussion}

\subsection{Comparison with Other Studies}  \label{sec:discuss2}
  In this section, we compare our results with other studies and discuss why the previous studies of low-redshift early-type galaxies found no or inverse (opposite to the theoretical expectation) environmental dependence.
 
 Some previous studies for low redshift were mostly limited to galaxies at $\log (M_{\star}/M_{\odot}) < 11.3$ and could not find the environmental dependence. For example, the sample of \citet{Cappellari2013} is limited to a mass range of $\log (M_{\star}/M_{\odot})\la11.3$ where the environmental dependence is negligible in our result. Similarly, the number of massive early-type galaxies at $\log (M_{\star}/M_{\odot}) > 11.3$ is very small in \citet{Maltby2010}, \citet{Nair2010}, and \citet{Poggianti2013}, all of which found no environmental dependence (see Table \ref{tabint} and its footnotes). 

  \citet{Cebrian2014} found that low-redshift field early-type galaxies are marginally larger than their counterparts in clusters by a factor of 1.05 or less (see their Figures 6 and 8). \citet{Poggianti2013} also found a similar trend  (see their Figures 11). This contradiction can be explained by the bias described in Section \ref{sec:Radius} that the sizes of galaxies in high-density environments can be slightly underestimated. However, figures of \citet{Cebrian2014} actually show that sizes of ``massive" early-type galaxies ($\log (M_{\star}/M_{\odot}) > 11.0$) in overdense environments are larger than those in underdense environments by a factor of $\sim 1.05$, which is in agreement with our result. 

\citet{Huertas2013b} claimed that there is no environmental dependence of mass--size relation for early-type galaxies. Interestingly, their mass--size relation for early-type galaxies (right panel of their Figure 1) shows a similar trend to that in \citet{Cebrian2014}. Early-type galaxies of $\log (M_{\star}/M_{\odot}) < 11.5$ in low-mass halos are marginally larger than their counterparts in high-mass halos.  However, this trend is reversed for massive early-type galaxies ($\log (M_{\star}/M_{\odot}) > 11.5$) in such a way that galaxies in high-mass halos are marginally larger than their counterparts in low-mass halos, which is consistent with our results.  
\\

\begin{figure}
\includegraphics[scale=0.147,angle=00]{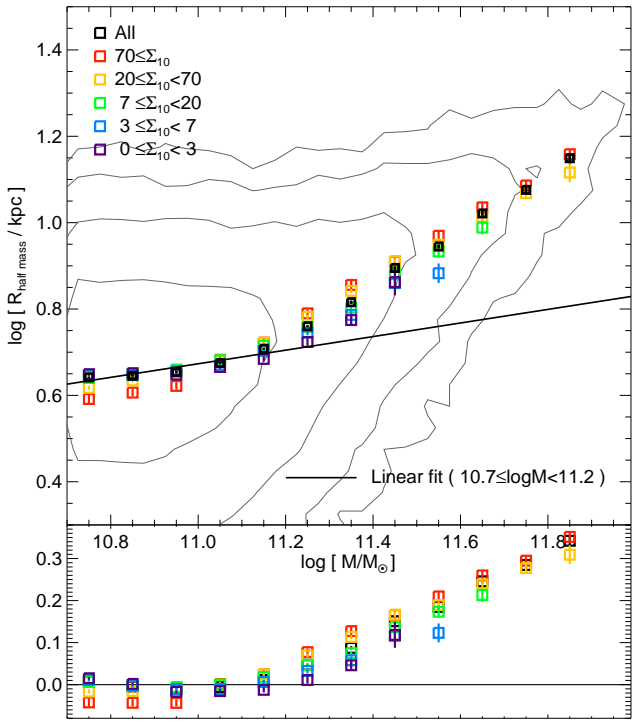}
\centering
	\caption{Mass--size relation of the simulated early-type galaxies from the $\Lambda$CDM galaxy formation simulation of \citet{Guo2011} based on the Millennium 1 Simulation \citep{Springel2005}. All symbols are the same as those in Figure \ref{ms_1fig}. 		\label{simul}}
\end{figure}

\begin{figure*}
\includegraphics[scale=0.15,angle=00]{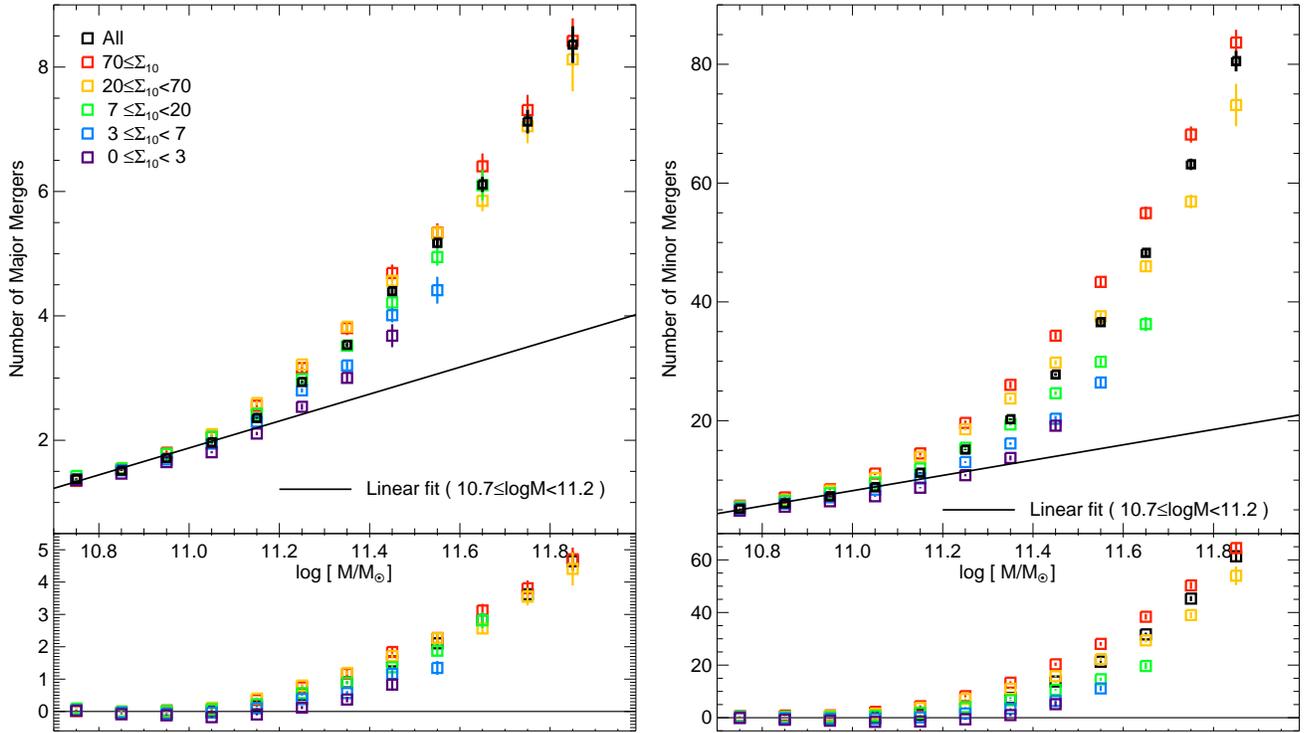}
\centering
	\caption{Average number of major mergers (left) and minor mergers (right) that the simulated galaxies have undergone, as a function of mass. The black squares indicate all galaxies, while the colored squares denote the galaxies under different environments. The error bar of each square point is $1\sigma$ of the average values from 200 bootstrap resampling. The solid black line is the best-fit relation for galaxies at $10.7\leq \log (M_{\star}/M_{\odot})<11.2$. Differences between all squares and the linear fit line are shown in the bottom subpanel. There is a clear environmental dependence of the merger history in the most mass galaxies of $\log (M_{\star}/M_{\odot})\ga11.0$. Below this mass range, the environmental dependence nearly disappears.
		\label{simul2}}
\end{figure*}

\begin{figure*}
\includegraphics[scale=0.15,angle=00]{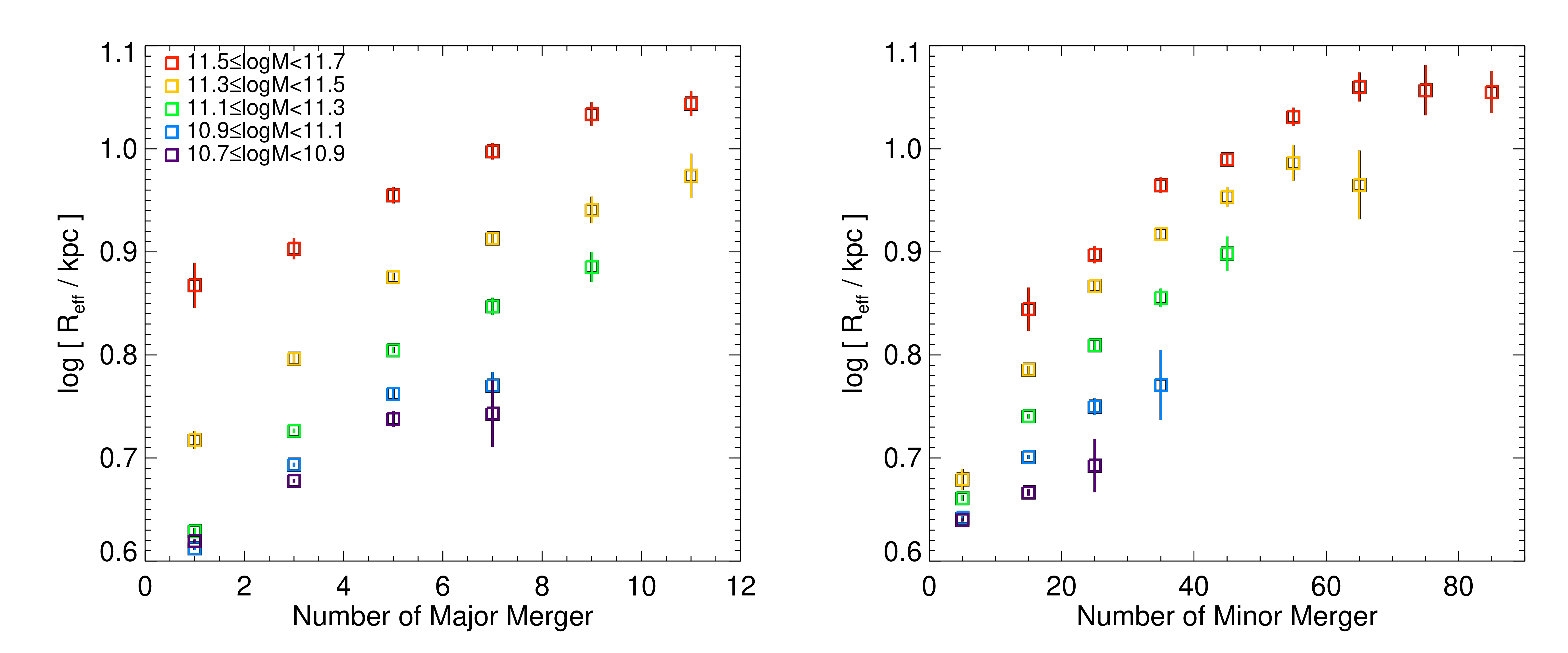}
\centering
	\caption{Final sizes of the simulated early-type galaxies as a function of the number of the major (left) and minor (right) mergers that the galaxies have undergone. The colored squares are median effective radii of the galaxies at different mass bins. The error bar of each square point is $1\sigma$ of the average effective radii of 200 bootstrap resampling. The early-type galaxies that have experienced more (major/minor) mergers have larger sizes than the counterparts that have experienced fewer mergers.
		\label{simul3}}
\end{figure*}

\subsection{Comparision with Simulation: Origin of Environmental Dependence}  \label{sec:discuss3}
  For comparision and interpretation, we used the $\Lambda$CDM galaxy formation simulation of \citet{Guo2011}\footnote{In this Section, we used the cosmological parameters used in \citet{Guo2011}.} based on the Millennium 1 Simulation \citep{Springel2005}. We selected early-type galaxies by a criterion of the bulge-to-total mass ratio (B/T) being larger than 0.7. We selected galaxies of $\log (M_{\star}/M_{\odot})>10.68$ in the redshift snapshot of 0.12 to match with the observational data we used. In the simulation, $96\%$ of the early-type galaxies selected by this method are found to have $\mathrm{B/T}>0.98$. For this reason, we regard the bulge mass as the mass of the galaxy and the bulge half-mass radius as the size of the galaxy. Note that the B/T distribution in simulation is different from the B/T distribution of the observed galaxies that extends to B/T $\sim 0.4$. However, as we mentioned in Appendix \ref{Appendix_A}, the observed low B/T value of the massive early-type galaxies is merely due to a mathematical formality, not due to the true existence of a significant disk component. Therefore, the difference in the B/T values does not stand as a significant problem in the comparison. The total number of early-type galaxies we used is 97,274.

 We also examined the color distribution of the simulated galaxies (see Appendix \ref{Appendix_A}). The color distribution is similar to the observed one, but not in perfect agreement. Because of the subtle difference in the galaxy properties between the simulation and the observation, we concentrate more on the qualitative aspect of the comparison between the observation and the simulation. 

 Environments were measured in the same way as what we used for the observational data. To make this possible, we converted the 3D coordinates of the simulation box to spherical coordinates with redshifts. We assumed the center of the snapshot box to be at a redshift of 0.12 and calculated distances between every galaxy and an observer at $z=0$. These distances were converted to redshifts of the simulated galaxies. In consideration of the fact that we used photometric redshifts, we scattered the redshifts randomly by adding random Gaussian error with $\sigma=\Delta z/(1+z)=0.022$ which corresponds to typical rms values of the difference between photometric redshifts and spectroscopic redshifts (6500km s$^{-1}$) in the observational data. We also performed a similar analysis in the 3D space (i.e., 3D local density). Not surprisingly, the environmental dependence becomes stronger in such an analysis, with the size difference between low- and high-density environments at 0.2 dex or so at a given mass. 

 Figure \ref{simul} shows the mass--size relation of the simulated early-type galaxies in different environments. There is an obvious environmental dependence at $\log (M_{\star}/M_{\odot})>11.2$, with the sizes in the high-density environment bigger than those in the low-density environment by 0.1 dex or so at a given mass. Interestingly, this is the exactly identical trend to the observational result shown in Figure \ref{ms_1fig}, despite the fact that the overall properties of the simulated galaxies do not match perfectly the properties of observed ones. Semianalytic galaxy simulation of \citet{Guo2011} can qualitatively reproduce the environmental dependence of the mass--size relation in the most massive galaxies. 

\begin{figure*}
\includegraphics[scale=0.15,angle=00]{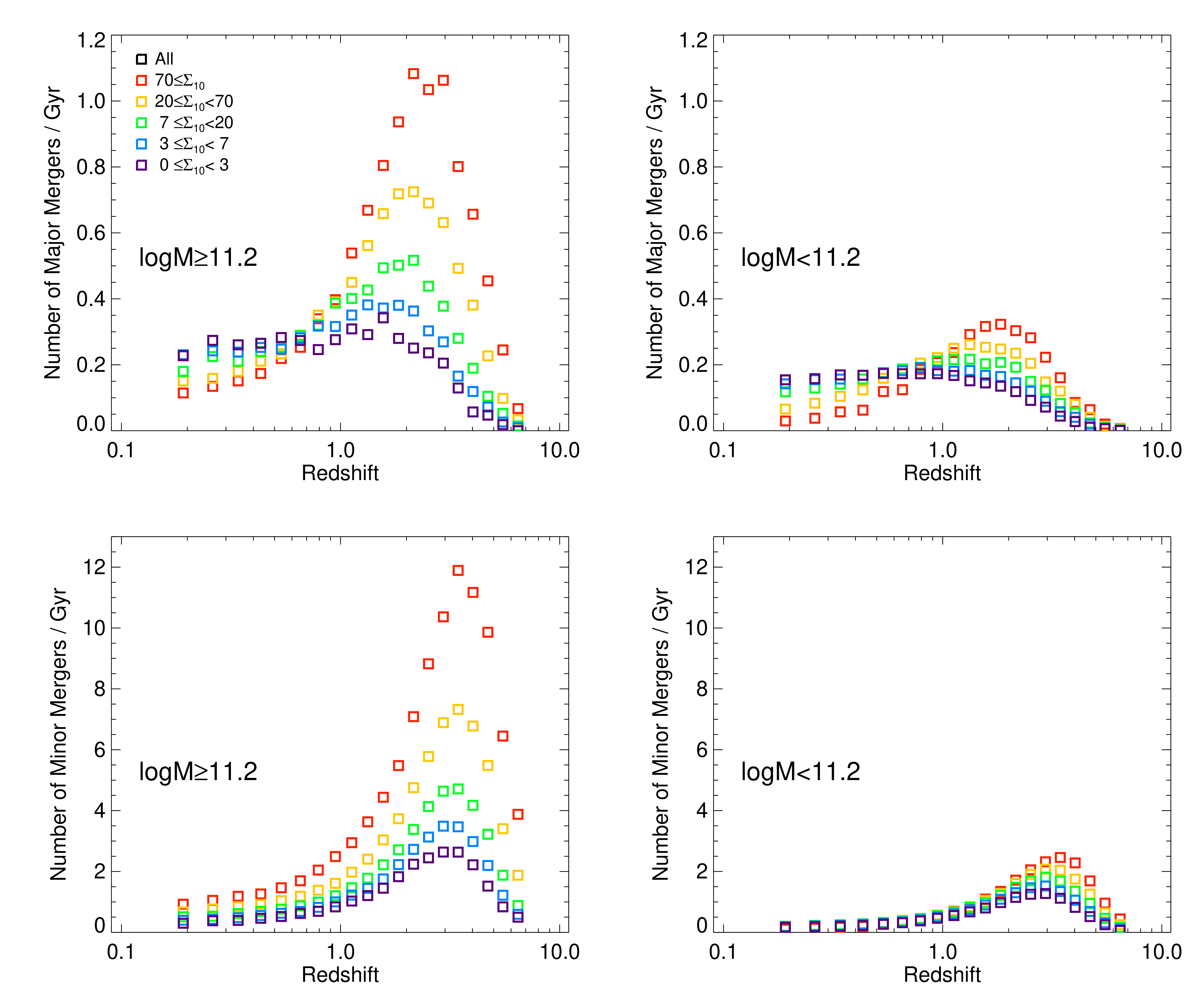}
\centering
	\caption{Look-back time distributions of major (upper panels) and minor (lower panels) merger events. The distributions for massive galaxies of $\log (M_{\star}/M_{\odot})\ge11.2$ are shown in the left panels, while those for less massive galaxies of $\log (M_{\star}/M_{\odot})<11.2$ are shown in the right panels. The colored squares are the average number of mergers per gigayear for different environments.
		\label{simul4}}
\end{figure*}

 There are some discrepancies between the observation and the simulation in the mass--size relation. The mass--size relation of the simulated galaxies has a very shallow slope of $\sim0.15$ in the low mass range of $10.7\leq \log (M_{\star}/M_{\odot})<11.2$, in contrast to 0.62 of the observed slope. This discrepancy might originate from the absence of a gas dissipation effect during gas-rich mergers in the simulation as stated in \citet{Guo2011}. In the gas-rich merger, angular momentum of gas is removed by gravitational torque and the gas falls toward the center of the galaxy \citep{Barnes1991,Barnes1996}, in which a starburst occurs, making a very compact central surface brightness profile \citep{Mihos1994,Mihos1996,Hopkins2008a,Hopkins2009a,Hopkins2009c} with a young, metal-rich population \citep{Lauer2005,Kuntschner2006,McDermid2006}, which is identified in several observations \citep[e.g.,][]{Hinkley2001,Im2001,Rothberg2004,Lauer2007b,Kormendy2009,KimIm2013}. The absence of this dissipation effect can cause larger sizes for simulated galaxies particularly in lower mass ranges, since their progenitors are likely to have larger gas fractions than those of massive galaxies. The large gas fractions for the low-mass progenitors can be inferred by the gas content of observed disk galaxies in $z\sim0$--$2$ \citep{Bell2000,Kannappan2004,McGaugh2005,Shapley2005,Erb2006}. Likewise, the simulation shows that the environmental trend is reversed for lower-mass galaxies in comparison to what we found for our sample of massive early-type galaxies. This is probably due to a similar reason to the very shallow slope in the mass--size relation that is explained above. This trend needs a further investigation. Another discrepancy is a larger scatter of the relation represented by the broad contour in Figure \ref{simul}. \citet{Guo2011} mention that this is a problem to be treated in more detail.

There are three methods of bulge growth in \citet{Guo2011}: major merger, minor merger, and disk instability. Here we briefly describe how the three methods work in the simulation. In the simulation, major merger combines all the stars of the progenitors to create a bulge component. Newly generated stars by cold gas during the merger are also assumed to reside in the bulge. Minor merger is treated somewhat differently. The bulge of the larger galaxy is assumed to gather all the stars of the smaller progenitor, whereas the newly formed stars are assigned to the disk component of the larger galaxy. For the mass and size growth of the bulge by the mergers, they used an energy conservation and a virial theorem as described by the equation (33) in \citet{Guo2011}. Bulge is also allowed to grow by a secular evolution in their simulation. When self-gravity of the disk is stronger than that of the dark matter halo in which the disk resides, the disk is considered to be unstable and the mass of the inner part of the disk is transferred to the bulge. 

 In order to understand the environmental dependence of the mass--size relation, we analyzed merger histories of all the simulated galaxies we selected and counted the number of major mergers and minor mergers that each galaxy has experienced.
 We classify a merger as a major merger when progenitors are more massive than $10^9\,M_{\odot}$ and the mass ratios of the progenitors are larger than 0.3 ($\mu>0.3$). On the other hand, a minor merger is defined as a merger when progenitors are more massive than $10^8\,M_{\odot}$ and the mass ratios of them are in the range of $0.01\le\mu\le0.3$. The mass limits are imposed, since mergers between low-mass objects will not affect the final size of massive galaxies much.

 Figure \ref{simul2} shows the average number of major mergers and minor mergers that the simulated galaxies have undergone, as a function of their stellar mass at $z=0.12$. We divided the galaxies into the same environment bins as in Figure \ref{simul}. There is a strong environmental dependence of the merger history in the most massive galaxies of $\log (M_{\star}/M_{\odot})\ga11.0$. Below this mass range, the environmental dependence nearly disappears. This trend is strikingly similar to the mass--size relation of the simulated galaxies shown in Figure \ref{simul} and that of the observed galaxies shown in Figure \ref{ms_1fig}. More massive galaxies have experienced more mergers. Moreover, massive galaxies in high-density environments have experienced more mergers than those in low-density environments. 

 The galaxies of $\log (M_{\star}/M_{\odot})\la11.0$ experience roughly one to two major mergers in their evolution history regardless of their environments. However, more massive galaxies, particularly in dense environments, underwent additional major mergers possibly with the galaxies already formed by previous mergers, and they also experience a very large number of minor mergers. 

 Figure \ref{simul3} shows final sizes of the simulated early-type galaxies as a function of the number of the major and minor mergers that the galaxies have undergone. It is shown that for a given mass, early-type galaxies that have experienced more (major/minor) mergers have larger sizes than the counterparts that have experienced fewer mergers. Dissipationless major mergers puff up the surface brightness profile and scatter the stars, making the profiles broader \citep{Hopkins2009b,Kormendy2009}. Minor mergers puff up the sizes of the galaxies very effectively, making the extended outer surface brightness profiles \citep{Bernardi2011a,Oogi2013}. Therefore, galaxies at a given mass in the simulation are larger if they have undergone more merger events. 

Additionally, we examined when the merger events happened in each environment. Figure \ref{simul4} shows look-back time distributions of major and minor merger events. For massive galaxies of $\log (M_{\star}/M_{\odot})\ge11.2$ in dense environments, the major merger rate has a peak at $z\sim2$ and declines thereafter. As the density of the environment goes down, the peak of the major merger rate moves to low redshift, and the major merger rate at the peak decreases substantially. At the low redshift, the trend is reversed in such a way that the massive galaxies in low-density environments have higher major merger rates. The distribution of minor merger events shows a different trend. The peak of the distribution is at $z\sim3$, and the peak moves a little to low redshift as the density of the environment goes down, compared with the case of the major merger. The minor merger rates decrease in all environments at low redshift without reversal of the merger rates. Less massive galaxies of $\log (M_{\star}/M_{\odot})<11.2$ show similar trend to the massive ones, except that their merger rate and its environmental dependence are lower than the massive ones. 
 
 The look-back time distribution of major merger events of the simulated early-type galaxies implies more rapid evolution in higher-density environments. The early-type galaxies in denser environments grew intensively at earlier ($z\sim2$) time than those in underdense environments. At low redshift, major merger events hardly occur in such a high-density environment, since the encounter velocities of galaxies are too high for major mergers to occur effectively \citep{Hopkins2008b}. On the other hand, the early-type galaxies in low-density environments merged and grew recently at moderate rates. 

In this section, we analyzed the merger histories of early-type galaxies in a simulation. The simulation results suggest that massive early-type galaxies in dense environments are larger than those in less dense environments. The environmental dependence in mass--size relation is qualitatively nearly identical to the observed result in Section \ref{sec:ms}, giving support for the galaxy formation in $\Lambda$CDM cosmological models. 
\\

\begin{figure*}
\includegraphics[scale=0.74,angle=00]{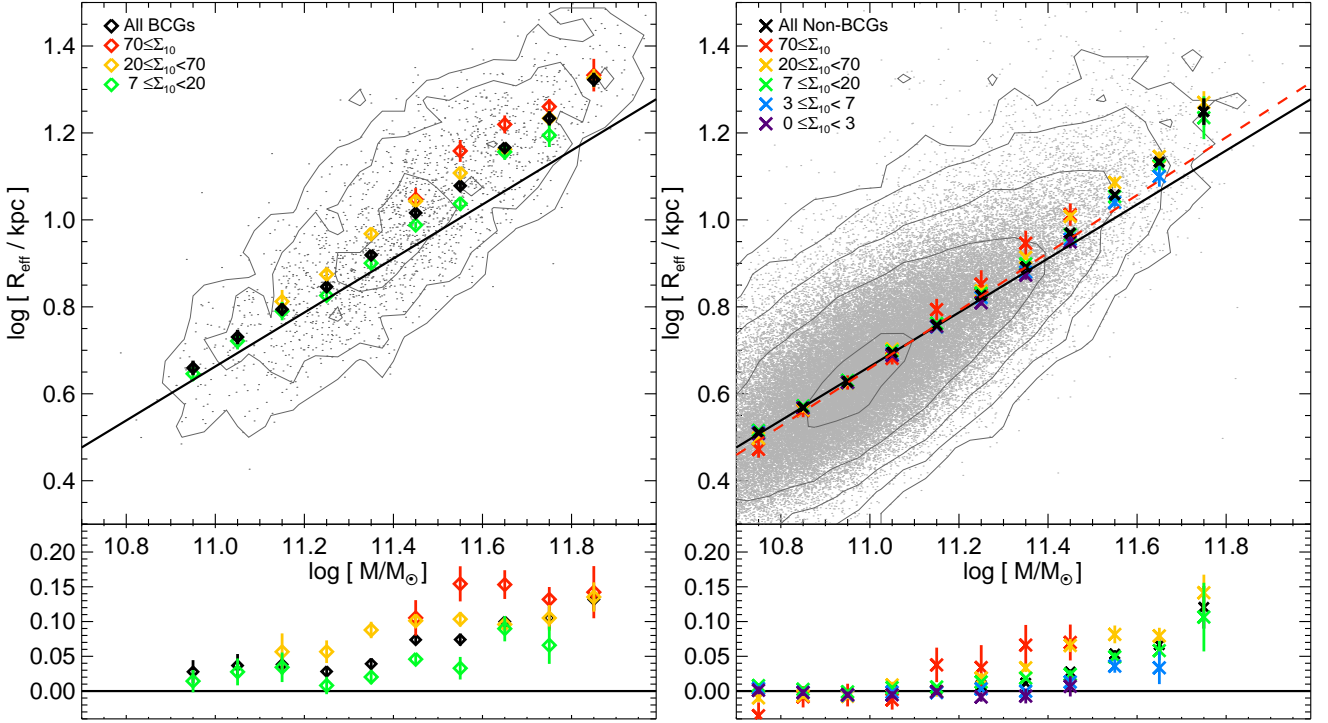}
\centering
	\caption{Mass--size relation of the BCGs (left panel) and non-BCGs (right panel). The solid black lines are the same line as that in Figure \ref{ms_1fig} for comparison.
		\label{ms_bcg_1fig}}
\end{figure*}

\begin{figure}
\includegraphics[scale=0.74,angle=00]{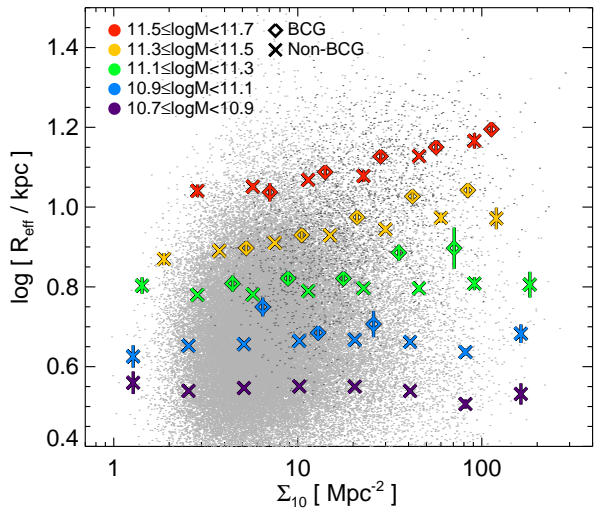}
\centering
	\caption{Environmental dependence of the effective radius in the size--density plane for BCGs and non-BCGs. The gray dots in background denote the non-BCGs, while the dark gray dots indicate the BCGs. The diamonds are median effective radii for the BCGs, whereas the crosses are those for the non-BCGs. 
		\label{ms_bcg_3fig}}
\end{figure}

\begin{deluxetable*}{ccccc}
\tablecaption{Median Effective Radii of BCGs [$\log(R_\mathrm{eff}/\mathrm{kpc})$]} 
\tabletypesize{\scriptsize}
\tablehead{
\colhead{Stellar Mass} & \colhead{All BCGs} & \colhead{$7\leq\Sigma_{10}<20$} & \colhead{$20\leq\Sigma_{10}<70$} & \colhead{$\Sigma_{10}\geq70$}
}
\startdata
$10.9\leq \log (M_{\star}/M_{\odot})<11.0$&$0.659\pm0.017$&$0.646\pm0.016$&\nodata&\nodata \\
$11.0\leq \log (M_{\star}/M_{\odot})<11.1$&$0.731\pm0.017$&$0.721\pm0.019$&\nodata&\nodata \\
$11.1\leq \log (M_{\star}/M_{\odot})<11.2$&$0.794\pm0.014$&$0.790\pm0.021$&$0.812\pm0.026$&\nodata \\
$11.2\leq \log (M_{\star}/M_{\odot})<11.3$&$0.846\pm0.008$&$0.826\pm0.013$&$0.875\pm0.016$&\nodata \\
$11.3\leq \log (M_{\star}/M_{\odot})<11.4$&$0.919\pm0.008$&$0.900\pm0.010$&$0.968\pm0.012$&\nodata \\
$11.4\leq \log (M_{\star}/M_{\odot})<11.5$&$1.016\pm0.006$&$0.988\pm0.007$&$1.043\pm0.009$&$1.048\pm0.026$ \\
$11.5\leq \log (M_{\star}/M_{\odot})<11.6$&$1.078\pm0.009$&$1.037\pm0.016$&$1.108\pm0.011$&$1.159\pm0.025$ \\
$11.6\leq \log (M_{\star}/M_{\odot})<11.7$&$1.166\pm0.006$&$1.156\pm0.018$&$1.162\pm0.007$&$1.220\pm0.021$ \\
$11.7\leq \log (M_{\star}/M_{\odot})<11.8$&$1.234\pm0.016$&$1.194\pm0.027$&$1.234\pm0.019$&$1.260\pm0.018$ \\
$11.8\leq \log (M_{\star}/M_{\odot})<11.9$&$1.322\pm0.016$&\nodata&$1.326\pm0.021$&$1.333\pm0.038$ \\
\enddata
\tablecomments{The unit of surface galaxy density $\Sigma_{10}$ is $\mathrm{Mpc}^{-2}$. The errors are $1\sigma$ of the median effective radii from 200 bootstrap resampling.}
\label{tab2}
\end{deluxetable*}

\begin{deluxetable*}{ccccccc}
\tablecaption{Median Effective Radii of Non-BCGs [$\log(R_\mathrm{eff}/\mathrm{kpc})$]} 
\tabletypesize{\scriptsize}
\tablehead{
\colhead{Stellar Mass} & \colhead{All Non-BCGs} & \colhead{$0\leq\Sigma_{10}<3$} & \colhead{$3\leq\Sigma_{10}<7$} & \colhead{$7\leq\Sigma_{10}<20$} & \colhead{$20\leq\Sigma_{10}<70$} & \colhead{$\Sigma_{10}\geq70$}
}
\startdata
$10.7\leq \log (M_{\star}/M_{\odot})<10.8$&$0.511\pm0.001$&$0.509\pm0.006$&$0.511\pm0.002$&$0.515\pm0.003$&$0.498\pm0.006$&$0.472\pm0.019$ \\
$10.8\leq \log (M_{\star}/M_{\odot})<10.9$&$0.568\pm0.001$&$0.567\pm0.005$&$0.568\pm0.002$&$0.572\pm0.002$&$0.563\pm0.005$&$0.561\pm0.015$ \\
$10.9\leq \log (M_{\star}/M_{\odot})<11.0$&$0.627\pm0.001$&$0.626\pm0.004$&$0.626\pm0.002$&$0.631\pm0.002$&$0.624\pm0.004$&$0.626\pm0.016$ \\
$11.0\leq \log (M_{\star}/M_{\odot})<11.1$&$0.695\pm0.001$&$0.688\pm0.005$&$0.692\pm0.002$&$0.699\pm0.002$&$0.703\pm0.005$&$0.681\pm0.014$ \\
$11.1\leq \log (M_{\star}/M_{\odot})<11.2$&$0.758\pm0.001$&$0.755\pm0.005$&$0.754\pm0.002$&$0.762\pm0.002$&$0.755\pm0.007$&$0.794\pm0.025$ \\
$11.2\leq \log (M_{\star}/M_{\odot})<11.3$&$0.826\pm0.002$&$0.809\pm0.007$&$0.821\pm0.003$&$0.831\pm0.002$&$0.837\pm0.007$&$0.852\pm0.032$ \\
$11.3\leq \log (M_{\star}/M_{\odot})<11.4$&$0.892\pm0.003$&$0.873\pm0.011$&$0.880\pm0.003$&$0.899\pm0.003$&$0.914\pm0.006$&$0.946\pm0.029$ \\
$11.4\leq \log (M_{\star}/M_{\odot})<11.5$&$0.970\pm0.004$&$0.949\pm0.015$&$0.955\pm0.006$&$0.966\pm0.006$&$1.008\pm0.009$&$1.012\pm0.026$ \\
$11.5\leq \log (M_{\star}/M_{\odot})<11.6$&$1.058\pm0.006$&\nodata&$1.040\pm0.010$&$1.053\pm0.006$&$1.086\pm0.013$&\nodata \\
$11.6\leq \log (M_{\star}/M_{\odot})<11.7$&$1.133\pm0.008$&\nodata&$1.100\pm0.023$&$1.125\pm0.014$&$1.146\pm0.012$&\nodata \\
$11.7\leq \log (M_{\star}/M_{\odot})<11.8$&$1.249\pm0.031$&\nodata&\nodata&$1.235\pm0.049$&$1.270\pm0.026$&\nodata \\
\enddata
\tablecomments{The unit of surface galaxy density $\Sigma_{10}$ is $\mathrm{Mpc}^{-2}$. The errors are $1\sigma$ of the median effective radii from 200 bootstrap resampling.}
\label{tab3}
\end{deluxetable*}

\subsection{Mass--Size Relation of Brightest Cluster Galaxies (BCGs)}  \label{sec:BCG}
The BCGs are massive early-type galaxies located at the center of clusters. Previous studies show that the BCGs have a different mass--size relation compared to non-BCGs \citep{Bernardi2007,Desroches2007,Lauer2007a,vonderLinden2007,Liu2008,Bernardi2009}. They show that BCGs have larger sizes and their mass--size relation is steeper than non-BCGs. Here we divide the sample into BCGs and non-BCGs, in order to study whether the environmental dependence of the mass--size relation in massive early-type galaxies is simply caused by an increasing fraction of BCGs in the massive galaxies in dense environments.

 To classify BCGs, we used the cluster catalog of \citet{Wen2012}. They identified 132,684 clusters and BCGs from SDSS-III using photometric redshift galaxies. The detection rate for clusters of $M_{200}\footnote{Cluster mass in $r_{200}$, where $r_{200}$ is the radius within which the mean density is 200 times the critical density of the universe.} > 1\times10^{14}\,M_{\odot}$ in the redshift range $0.05\leq z < 0.42$  is above $95\%$, and the false detection rate for the whole sample is less than $6\%$. This classification is also well matched with other previous studies,\footnote{Abell sample from the Palomar Sky Survey \citep{Abell1989}, maxBCG \citep{Koester2007}, WHL09 \citep{Wen2009}, GMBCG \citep{Hao2010}, and AMF cluster sample \citep{Szabo2011}} although it depends on the richness of clusters ($50\%$ -- $70\%$, but 90\% for rich clusters). The number of BCGs classified by the catalog of \citet{Wen2012} is 1897. At the massive end ($\log (M_{\star}/M_{\odot})>11.6$), 52\% of early-type galaxies (410/788) are classified as BCGs.

 Figure \ref{ms_bcg_1fig} shows the mass--size relation of the BCGs and the non-BCGs. The diamonds are the median effective radii for the BCGs, whereas the crosses are those for the non-BCGs. The mass--size relations of the BCGs and the non-BCGs are summarized in Tables \ref{tab2} and \ref{tab3}, respectively. Figure \ref{ms_bcg_3fig} shows the environmental dependence of their sizes in the size--density plane.

 The non-BCGs show a similar mass--size relation to the case for full sample. We found that early-type non-BCGs more massive than $\log (M_{\star}/M_{\odot})\sim11.2$ show also environmental dependence in the mass--size relation. When the BCGs are removed from the full sample, the size differences between high density environments and low-density environments at $\log (M_{\star}/M_{\odot})>11.2$ are 0.00 -- 0.02 dex smaller than before, but the environmental dependence of the sizes does not disappear.

 Most of the BCGs are more massive than $\log (M_{\star}/M_{\odot})\sim11.1$ and reside in environments denser than $\Sigma_{10}\geq7\,\mathrm{Mpc}^{-2}$. The median sizes of BCGs are larger than non-BCGs particularly in the mass range of $11.0< \log (M_{\star}/M_{\odot})<11.5$. The BCG population also shows environmental dependence in the mass--size relation. We constructed a mass-matched sample for BCGs in the same way as described in Appendix \ref{Appendix_B}, and tested whether the different mass distributions in a mass bin between different environments can cause the environmental dependence of the mass--size relation of BCGs. By doing so, we found that the size variation is very small ($\la0.01$ dex) as it is for the full sample, and conclude that the environmental dependence of the mass--size relation of BCGs is not driven by simple difference of mass distribution in a mass bin between different environments.

  Due to the increasing fraction of BCGs in the massive galaxies in dense environments, the environmental dependence of the mass--size relation for the full sample can be partially amplified \citep[e.g,][]{Zhao2015}. However, considering the facts that both BCGs and non-BCGs show environmental dependence of their mass--size relations in the massive galaxies and the non-BCGs show a similar trend of mass--size relation to the case for the full sample, the environmental dependence is not solely caused by the increasing fraction of the BCGs in the massive galaxies in dense environments. 

\citet{Liu2008} revealed that many BCGs have extended stellar envelopes in their outskirts, which is consistent with our result. \citet{Bernardi2007} suggested that the increasing fraction of BCGs causes the curved mass--size relation in the most massive galaxies and non-BCGs do not show curved relation. On the other hand, \citet{Desroches2007} and \citet{vonderLinden2007} showed that there is curvature in the mass--size relation not only for BCGs but also for non-BCGs. Our result supports the results of \citet{Desroches2007} and \citet[see Figure \ref{ms_bcg_1fig}]{vonderLinden2007}.

\citet{Zhao2015} showed that the environment does not affect sizes of BCGs. On the other hand, \citet{Garilli1997} found that BCGs with large effective radii are residing in a denser environment, which is consistent with our result. 
\\

\section{Summary}
 We investigated the mass--size relation of the most massive early-type galaxies in the local universe. We used 73,116 early-type galaxies in the redshift range of $0.1\le z<0.15$ from the spectroscopic sample of SDSS DR7. We measured the galaxy environments using all SDSS photometric objects classified as galaxies and their photometric redshifts. We defined the environment as the surface number density of galaxies within the 10th nearest neighbor, using the galaxies whose photometric redshifts fall between velocity cuts ($\pm6500$ km s$^{-1}$) centered on the spectroscopic redshifts of early-type galaxies. The galaxy sizes are measured by a nonparametric method, in which $r$-band images are used. We masked all other sources in the images and extracted curve of growth (cumulative flux) and calculated a radius where one-half of the total flux is contained. We corrected a seeing effect using a simple empirical relation.

 We tested our nonparametric method by using simulated galaxies generated by GALFIT software. From the test, we conclude that the measured sizes coincide with the input sizes of the various models of the simulated galaxies, and neighboring sources do not severely affect the size and total flux measurement. 

 We found environmental dependence of the mass--size relation in the most massive early-type galaxies of $\log (M_{\star}/M_{\odot})>11.2$. Massive galaxies in high-density environments have larger sizes than their counterparts in low-density environments. On the other hand, there is negligible environmental dependence for the galaxies in the mass range of $10.7\leq \log (M_{\star}/M_{\odot})<11.2$.

 We discussed why the previous studies of low-redshift early-type galaxies found no environmental dependence. Some previous studies for low redshift do not have enough galaxies at $\log (M_{\star}/M_{\odot}) > 11.2$, where we found the environmental dependence of the mass--size relation. Results of other studies show marginal environmental dependence in the massive galaxies at $\log (M_{\star}/M_{\odot}) > 11.2$, which is in agreement with our result.
 
 To interpret the result, we used $\Lambda$CDM galaxy formation simulation of \citet{Guo2011} based on the Millennium 1 Simulation \citep{Springel2005}. We found that this galaxy simulation can qualitatively reproduce the environmental dependence of the mass--size relation in the most massive galaxies, except some discrepancies. In order to understand the environmental dependence of the mass--size relation, we analyzed merger histories of all the simulated galaxies we selected, and we counted the number of the major mergers and minor mergers that each galaxy has experienced. The number of the mergers they experienced shows a clear environmental dependence in the most massive galaxies of $\log (M_{\star}/M_{\odot})\ga11.0$. Below this mass range, the environmental dependence nearly disappears. This trend is strikingly consistent not only with the mass--size relation of the simulated galaxies but also with that of the observed galaxies. We conclude based on the simulation data that massive galaxies underwent major and minor mergers more frequently in high-density environments, and this caused them to have larger sizes than their counterparts in low-density environments. 

  We divided the sample into BCGs and non-BCGs. The non-BCGs show a similar mass--size relation to the case for the full sample. The median sizes of the BCGs are larger than those of the non-BCGs in the mass range of $11.0< \log (M_{\star}/M_{\odot})<11.5$.  Both populations also show environmental dependence of their mass--size relations in the massive galaxies.

 The simulation results show that most of the major merger events took place at $z > 1$ for early-type galaxies in high-density environments, while it is so at $z < 1$ for early-type galaxies in less dense environments. Therefore, we expect to see a stronger environmental dependence as we go to higher redshifts. Previous results at $z \sim 1$ already show such a trend, but a more thorough analysis should tell us if the model predictions are correct or not.  
\\

\acknowledgments 
This work was supported by the National Research Foundation of Korea (NRF) grant no. 2008-0060544, funded by the Korea government (MSIP).
Funding for the SDSS and SDSS-II has been provided by the Alfred P. Sloan Foundation, the Participating Institutions, the National Science Foundation, the U.S. Department of Energy, the National Aeronautics and Space Administration, the Japanese Monbukagakusho, the Max Planck Society, and the Higher Education Funding Council for England. The SDSS Web site is http://www.sdss.org/.
The SDSS is managed by the Astrophysical Research Consortium for the Participating Institutions. The Participating Institutions are the American Museum of Natural History, Astrophysical Institute Potsdam, University of Basel, University of Cambridge, Case Western Reserve University, University of Chicago, Drexel University, Fermilab, the Institute for Advanced Study, the Japan Participation Group, Johns Hopkins University, the Joint Institute for Nuclear Astrophysics, the Kavli Institute for Particle Astrophysics and Cosmology, the Korean Scientist Group, the Chinese Academy of Sciences (LAMOST), Los Alamos National Laboratory, the Max-Planck-Institute for Astronomy (MPIA), the Max-Planck-Institute for Astrophysics (MPA), New Mexico State University, Ohio State University, University of Pittsburgh, University of Portsmouth, Princeton University, the United States Naval Observatory, and the University of Washington. 
The Millennium Simulation databases used in this paper and the web application providing online access to them were constructed as part of the activities of the German Astrophysical Virtual Observatory (GAVO).
\\

\appendix
\section{Characteristics of Our Sample of Early-Type Galaxies}  \label{Appendix_A}
\begin{figure}
\includegraphics[scale=0.17,angle=00]{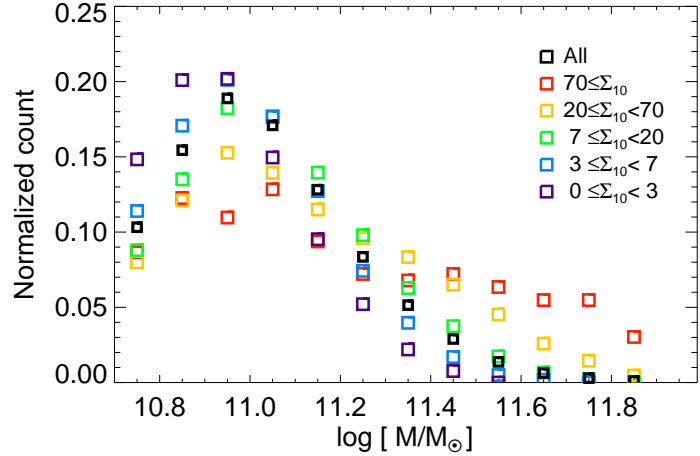}
\centering
	\caption{Stellar mass distribution divided by the environments. In this figure, each distribution is normalized by the total number of galaxies in each environment, respectively (see, Figure \ref{denhistfig}).
		\label{md_fig}}
\end{figure}

 We provide a stellar mass distribution of our sample of early-type galaxies divided by the environments in Figure \ref{md_fig}. In this figure, each distribution is normalized by the total number of galaxies in each environment, respectively (see, Figure \ref{denhistfig}). The stellar mass distributions are stretched to higher mass for higher-density environments. However, as we noted in Appendix \ref{Appendix_B}, the environmental dependence of the mass--size relation is not caused by simple difference of mass distribution between different environments.

\begin{figure}
\includegraphics[scale=0.25,angle=00]{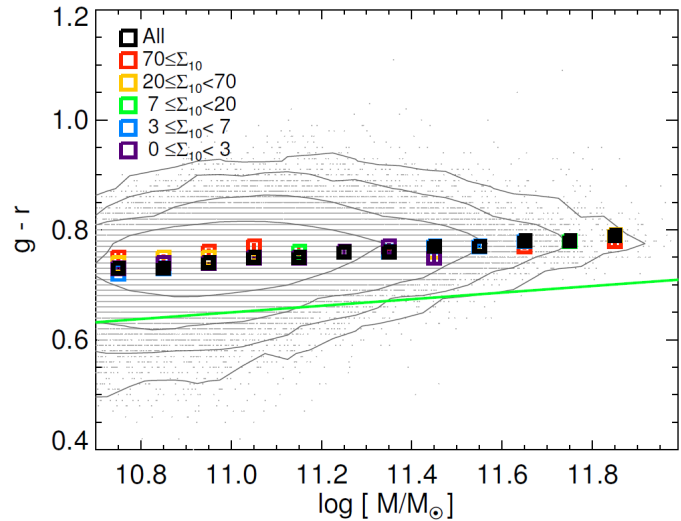}
\centering
	\caption{Rest-frame $g-r$ color--mass diagram for our sample of early-type galaxies. The squares indicate median values, and the error bar of each square is $1\sigma$ of the median values from 200 bootstrap resampling. The green line divides galaxies into red and blue populations as in \citet{Bluck2014}. Most of the galaxies are above the green line, indicating that they are red galaxies. Furthermore, their median color values are very similar ($0.7$--$0.8$) across the whole mass range in the figure and show negligible environmental dependence. 
		\label{cmd_fig}}
\end{figure}

Figure \ref{cmd_fig} shows the rest-frame $g-r$ color--mass diagram for our sample of early-type galaxies. The green line in this figure is the line dividing galaxies into red and blue populations as in \citet{Bluck2014}. Most of the galaxies are above the green line, indicating that they are red galaxies. Furthermore, their median color values are very similar ($0.7$--$0.8$) across the whole mass range in the figure and show negligible environmental dependence.

\begin{figure}
\includegraphics[scale=0.25,angle=00]{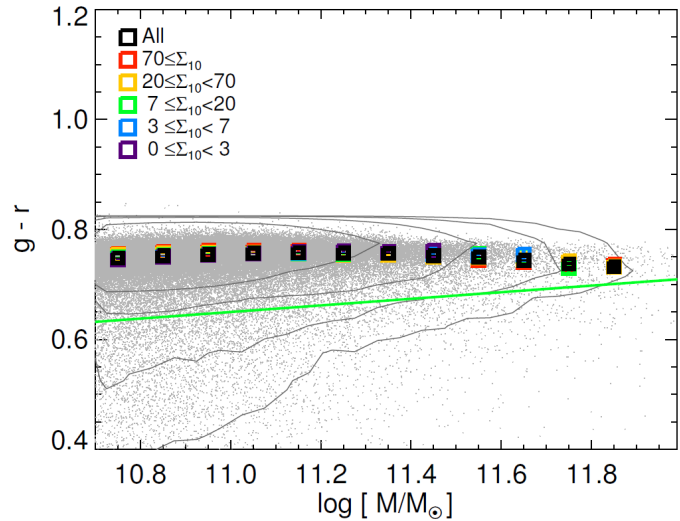}
\centering
	\caption{Rest-frame $g-r$ color--mass diagram for early-type galaxies from simulation. The squares indicate median values, and the error bar of each square is $1\sigma$ of the median values from 200 bootstrap resampling. The green line divides galaxies into red and blue populations as in \citet{Bluck2014}. The color distribution of the simulated sample shows a tail extended to blue color in comparison with that of the observational sample. However, the color distribution of the simulated sample is similar to that of the observational sample in the sense that most of the galaxies are above the green line, and their median color values are very similar ($0.7$--$0.8$) across the whole mass range without environmental dependence. 
		\label{cmd_simul_fig}}
\end{figure}

For comparison, we also present a rest frame $g-r$ color--mass diagram for early-type galaxies from simulation in Figure \ref{cmd_simul_fig}. The color distribution of the simulated sample shows a tail extended to blue color in comparison with that of the observational sample. However, the color distribution of the simulated sample is similar to that of the observational sample in the sense that most of the galaxies are above the green line, and their median color values are very similar ($0.7$--$0.8$) across the whole mass range without environmental dependence. 

\begin{figure}
\includegraphics[scale=0.17,angle=00]{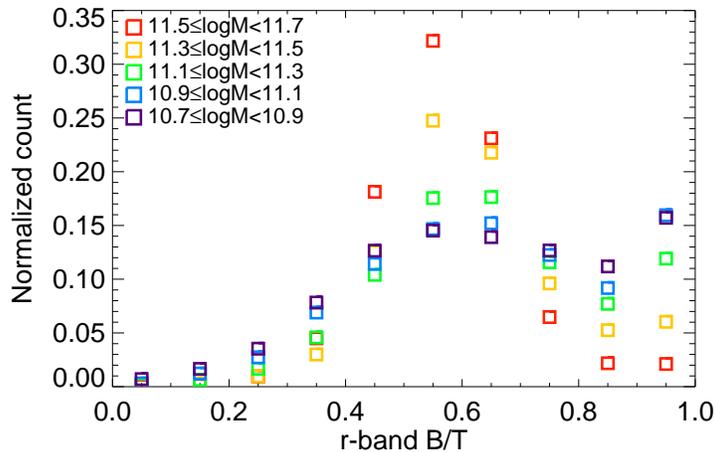}
\centering
	\caption{$r$-band B/T distribution of our sample extracted from S11. We divided them by several mass ranges. Each distribution is normalized by the total number of galaxies in each mass range, respectively. The distribution stretches down to B/T $\sim 0.4$ even in the most massive regime, but we note that the B/T presented in S11 is not an ideal indicator of ``genuine" B/T for all galaxies. In some cases, bulge and disk components are not real bulge and disk, but just a mathematical representation of the surface brightness profiles. See Appendix \ref{Appendix_A} for details. 
		\label{bt_fig}}
\end{figure}

Figure \ref{bt_fig} shows the $r$-band B/T distribution of our sample extracted from S11. We divided them by several mass ranges. The distribution stretches down to B/T $\sim 0.4$ even in the most massive regime, but we note that the B/T presented in S11 is not an ideal indicator of ``genuine" B/T for all galaxies, but just a mathematical representation of the surface brightness profiles. For example, massive galaxies ($\log (M_{\star}/M_{\odot}) > 11.3$) frequently have low B/T values of $\sim0.5$ in S11 and M14. But this does not indicate that the most massive early-type galaxies have ``real'' disks. As explained by M14, they are fitted to have a significant disk component due to extended outer surface brightness profiles. S11 also mentioned that disks in bulge-dominated galaxies are not genuine but caused by an additional degree of freedom to model the outer wings of galaxies.

\begin{figure}
\includegraphics[scale=0.25,angle=00]{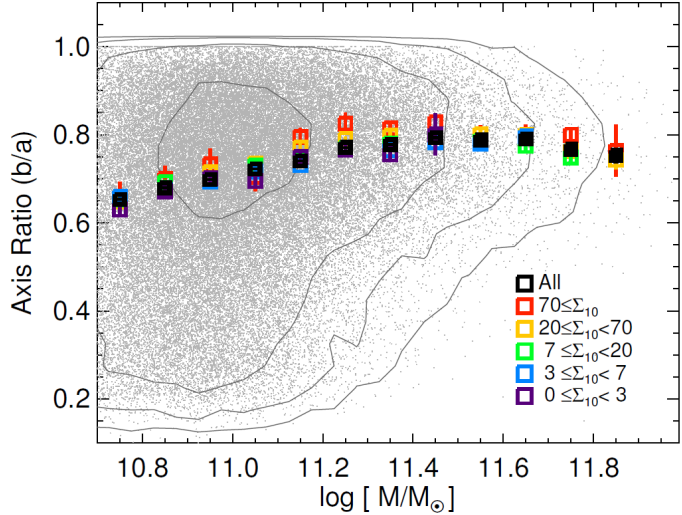}
\centering
	\caption{Axis ratios of our sample of early-type galaxies as a function of stellar mass. The squares indicate median values, and the error bar of each square is $1\sigma$ of the median values from 200 bootstrap resampling. The axis ratio distribution of our sample shows that median axis ratios of the galaxies in the highest-density environment are $\sim0.05$ larger than those in the low-density environments. This is probably due to the fact that early-type galaxies in the low-density environments have a bit more disk-like structure. However, environmental dependence of the axis ratios in our sample is small.
		\label{ba_fig}}
\end{figure}

 The axis ratio distribution of our sample presented in Figure \ref{ba_fig} shows that median axis ratios of the galaxies in the highest-density environment are $\sim0.05$ larger than those in the low-density environments. This is probably due to the fact that early-type galaxies in low-density environments have a bit more disk-like structure by the morphology-density relation \citep{Dressler1980,Houghton2015}, in which the population ratio of elliptical galaxies to lenticular galaxies is higher at the higher projected density environment. However, environmental dependence of the axis ratios in our sample is small.

\section{Mass--Size Relations Derived by Various Methods}  \label{Appendix_B}
 We show here mass--size relations of early-type galaxies analyzed in many different ways to demonstrate that our main conclusion of the environmental dependence is not affected by details of how the sample is analyzed.  
 
\begin{figure}
\includegraphics[scale=0.31,angle=00]{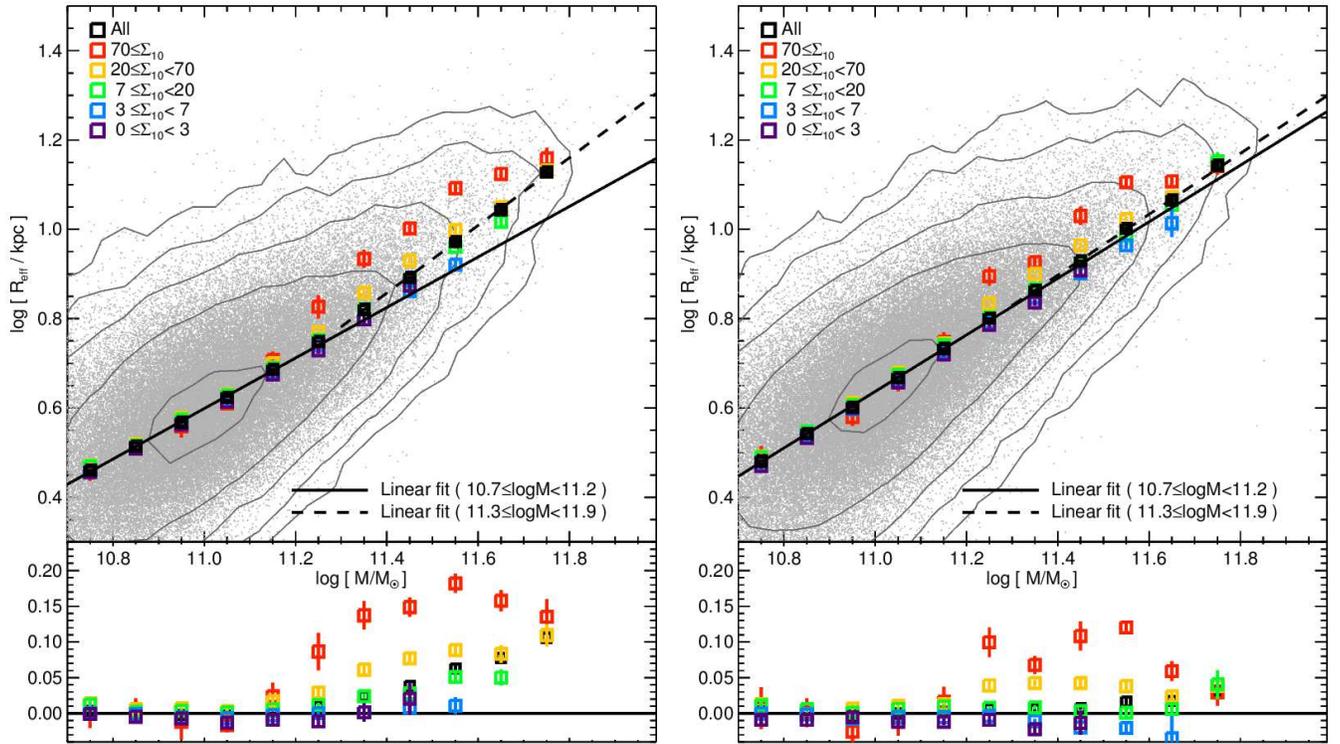}
\centering
	\caption{Mass--size relations of our full sample of early-type galaxies (79,184 early-type galaxies) based on effective radii and stellar masses derived by the de Vaucouleurs + exponential profile fit (left panel) and S\'{e}rsic + exponential profile fit (right panel) from the S11 and M14 catalogs. All symbols are the same as those in Figure \ref{ms_1fig}.
		\label{ms_simard}}
\end{figure}

 First, we show that the exclusion of the objects with close bright neighbors (see Section \ref{sec:Data}) does not influence our main conclusion, by presenting the mass--size relation of our full sample of early-type galaxies (79,184 early-type galaxies, including those that were excluded in the mass--size relation using nonparametric size measurements due to close neighbors) based on effective radii and stellar masses derived by the model fit from S11 and M14 catalogs. The mass--size relations based on the de Vaucouleurs + exponential profile fit is shown in the left panel of Figure \ref{ms_simard}, while that based on the S\'{e}rsic + exponential profile fit is shown in the right panel of Figure \ref{ms_simard}.\footnote{The M14 catalog does not give stellar masses for the S\'{e}rsic + exponential profile fit. Therefore, for this case we derived stellar masses using Equation \ref{eq3} where the magnitudes and color values are from the S\'{e}rsic + exponential profile fit.} In these two cases, there is also environmental dependence ($0.1$--$0.15$ dex) of the mass--size relation for the massive galaxies of $\log (M_{\star}/M_{\odot}) > 11.2$. We also found that slopes of the mass--size relations in the low mass range of $10.7\leq \log (M_{\star}/M_{\odot})<11.2$ are similar to what we found in Section \ref{sec:ms} with 0.563 for the de Vaucouleurs + exponential profile fit and 0.634 for the S\'{e}rsic + exponential profile fit. 

\begin{figure}
\includegraphics[scale=0.31,angle=00]{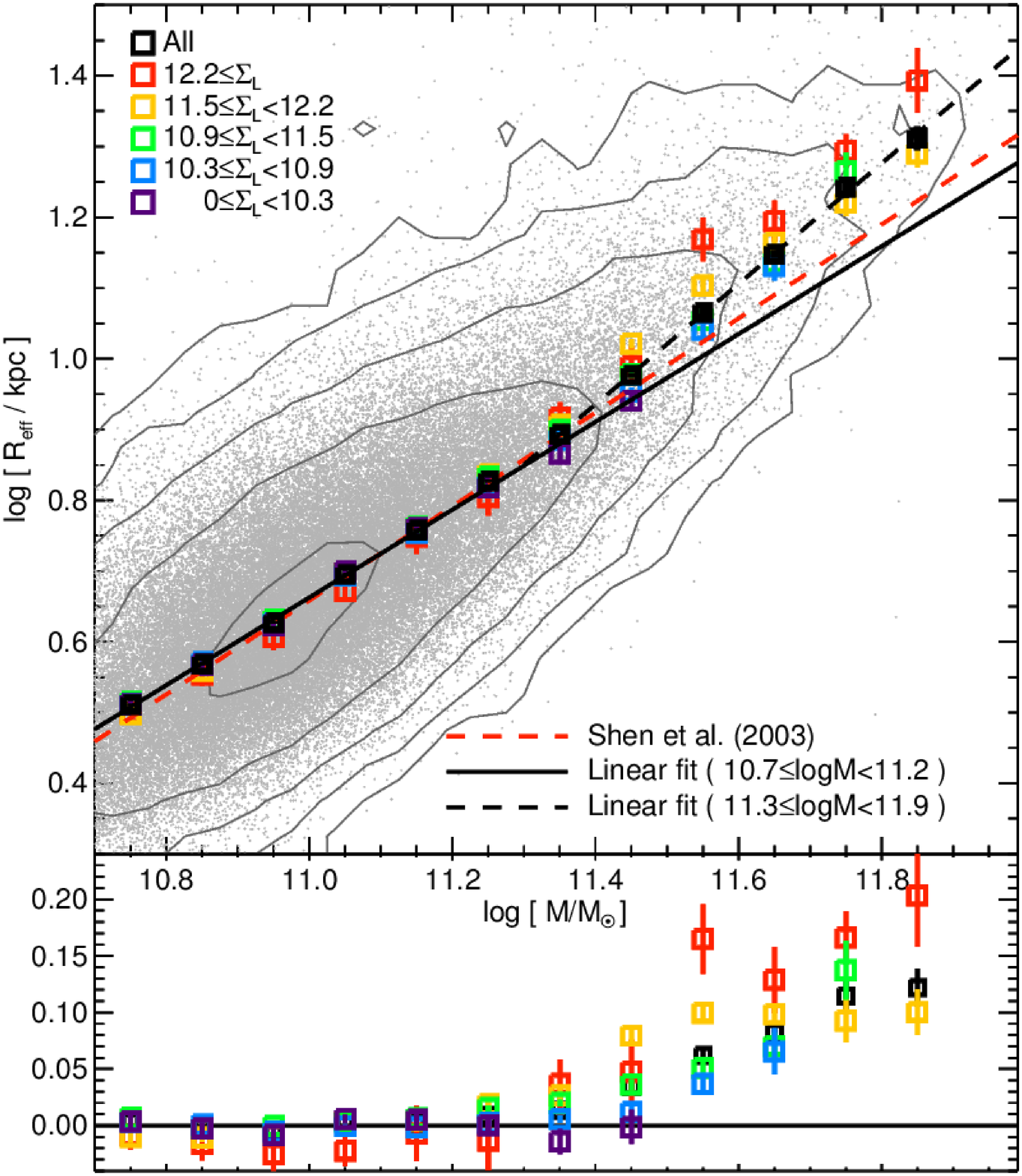}
\centering
	\caption{Mass--size relation of early-type galaxies when the environments are measured by $r$-band luminosity density within the 10th nearest objects. The unit of luminosity density $\Sigma_{L}$ is $L_{\odot}\,\mathrm{Mpc}^{-2}$. We divided galaxies into five environments so that each environment bin has a similar number of galaxies to that of each environment in Figure \ref{ms_1fig}. All symbols are the same as those in Figure \ref{ms_1fig}. 
		\label{ms_lumden}}
\end{figure}

\begin{figure}
\includegraphics[scale=0.31,angle=00]{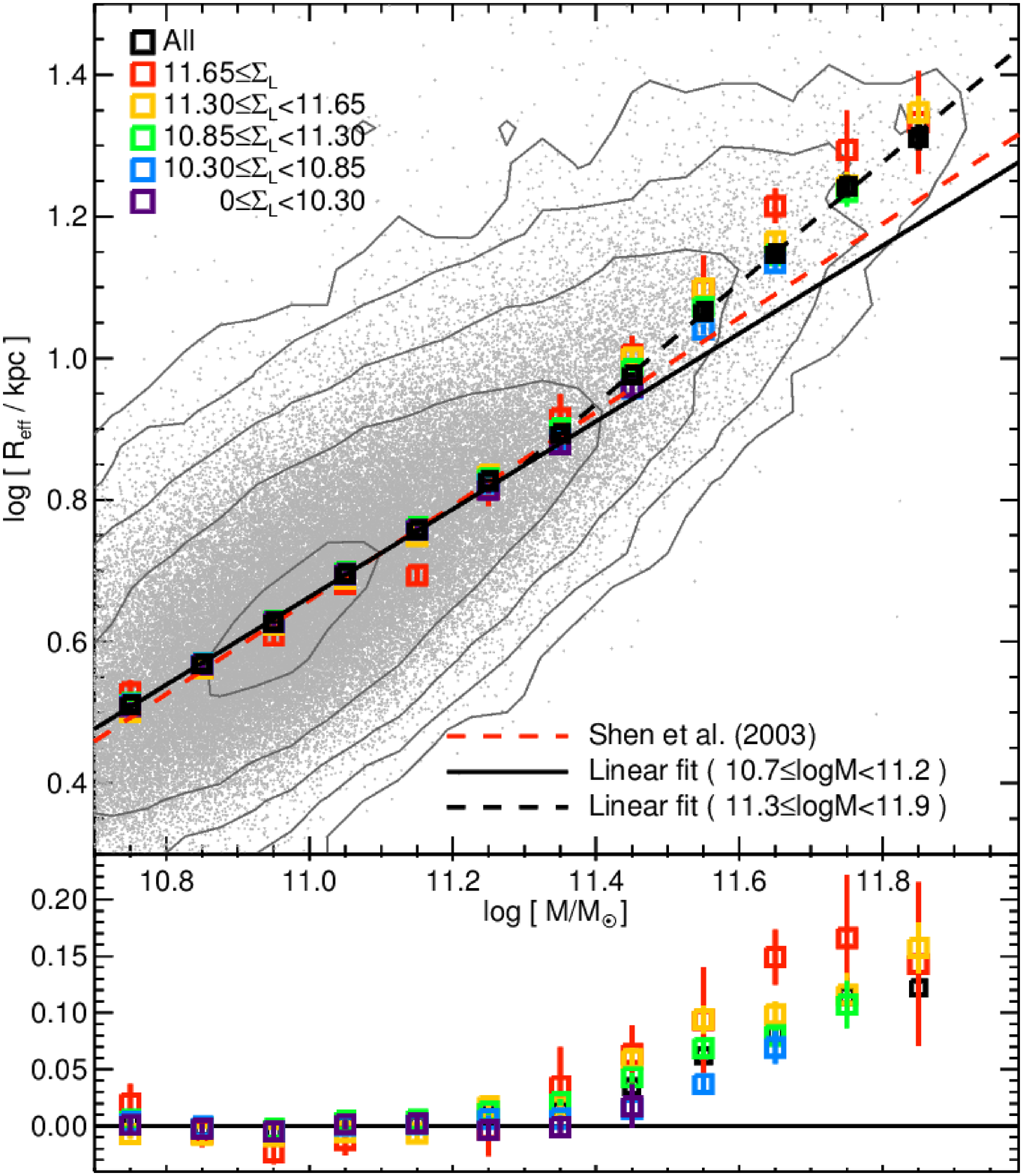}
\centering
	\caption{Mass--size relation of early-type galaxies when the environments are measured by $r$-band luminosity density within 1 Mpc. The unit of luminosity density $\Sigma_{L}$ is $L_{\odot}\,\mathrm{Mpc}^{-2}$. We divided galaxies into five environments so that each environment bin has a similar number of galaxies to that of each environment in Figure \ref{ms_1fig}. All symbols are the same as those in Figure \ref{ms_1fig}. 
		\label{ms_lumden_ap}}
\end{figure}

 We provide the mass--size relations of early-type galaxies when the environments are measured by $r$-band luminosity density within the 10th nearest objects in Figure \ref{ms_lumden} and within 1 Mpc in Figure \ref{ms_lumden_ap}. The unit of luminosity density $\Sigma_{L}$ is $L_{\odot}\,\mathrm{Mpc}^{-2}$. In these figures, we divided galaxies into five environments so that each environment bin has a similar number of galaxies to that of each environment in Figure \ref{ms_1fig}. Use of the luminosity density reaches essentially the same result.

\begin{figure}
\includegraphics[scale=0.31,angle=00]{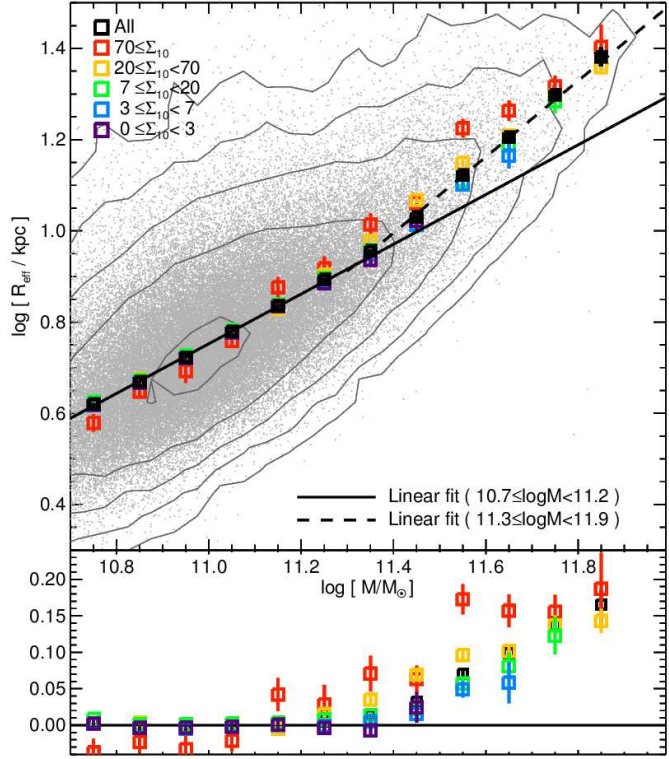}
\centering
	\caption{Mass--size relation of early-type galaxies when half-light semimajor axes are used. The half-light semimajor axes are calculated by dividing the half-light radii derived in Section \ref{sec:Radius} by square root values of the axis ratios from SDSS DR7. All symbols are the same as those in Figure \ref{ms_1fig}. Use of a half-light semimajor axis increases the sizes of the early-type galaxies, but it does not change our main conclusion.
		\label{ms_sma}}
\end{figure}

 We also provide the mass--size relation of early-type galaxies when half-light semimajor axes are used in Figure \ref{ms_sma}. The half-light semimajor axes were calculated by dividing the half-light radii derived in Section \ref{sec:Radius} by square root values of the seeing-corrected axis ratios from SDSS DR7. This is because our nonparametric method is not ideal for deriving the half-light semimajor axis, particularly for small galaxies, because the axis ratio of the elliptical aperture is affected by seeing, unlike a circular aperture. Use of a half-light semimajor axis increases the sizes of the early-type galaxies, but it does not affect the environmental dependence we find in the massive end. On the other hand, we find a weak environmental dependence in an opposite way at low-mass galaxies. The amount of the dependence is small ($\sim$ 0.03 dex) and comparable to the bias in size measurements in crowded environments, so it is not easy to judge whether this trend is real. It is worthwhile to examine this reverse environmental dependence further in a future work. 
 
 We also examined whether the environmental dependence is driven by a difference in the stellar mass distribution between different environments at a given mass bin, since stellar mass is coupled with environment in a sense that massive galaxies preferentially inhabit dense environments. Thus, for a given stellar mass bin, the stellar mass distribution of high-density environments could be skewed more toward massive galaxies in comparison to that of low-density environments. The different mass distribution in a mass bin between different environments can possibly bias the environmental dependence of the mass--size relation. To test this hypothesis, we constructed the mass-matched sample as follows: for each stellar mass bin in the high density environments ($20\leq\Sigma_{10}<70$ and $\Sigma_{10}\geq70$), we created a randomly selected sample whose mass distribution is the same as that of the counterpart in the lowest-density environments in that mass range (we repeated this random sampling 10,000 times for robustness). This was done from $\log (M_{\star}/M_{\odot})=11.3$ to the most massive bin where the environmental dependence exists. Then we examined how much the sizes of the galaxies in the high-density environments change and how much weaker the environmental dependence becomes. When the mass-matched sample described above is used, the median sizes of the galaxies in the high-density environments become on average $0.001$ dex (median difference of $0.000$ dex) smaller than before, with a standard deviation of $0.011$ dex. This small change is negligible when examining the environmental dependence, which is of the order of 0.1 dex.

 We repeated the same test in reverse: for each stellar mass bin in the low-density environments ($0\leq\Sigma_{10}<3$ and $3\leq\Sigma_{10}<7$), we created a randomly selected sample whose mass distribution is the same as that of the counterpart in the highest-density environments. In this case, the median sizes of the galaxies in the low-density environments become on average $0.003$ dex (median difference of $0.003$ dex) larger than before, with a standard deviation of $0.003$ dex. Again, the change is very negligible.

These tiny variations are smaller than typical errors ($0.01$ -- $0.02$ dex) of the median sizes of each bin at $\log (M_{\star}/M_{\odot})>11.3$ and do not change environmental dependence significantly. Therefore, we conclude that the environmental dependence of the mass--size relation we found is driven by a difference in the mass--size relation, not by a simple difference of mass distribution in a mass bin between different environments. 

\begin{figure}
\includegraphics[scale=0.74,angle=00]{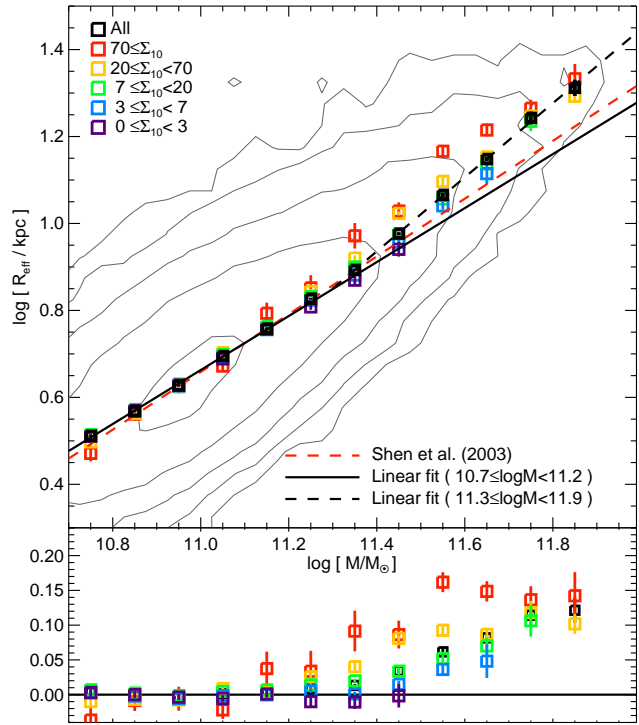}
\centering
	\caption{Mass--size relation of the mass-matched sample. Here we matched the mass distribution of each mass bin with that of the counterpart in the environment of $20\leq\Sigma_{10}<70$. All symbols are the same as those in Figure \ref{ms_1fig}. The use of the mass-matched sample does not change our result.
		\label{msmatched_fig}}
\end{figure}

\begin{figure}
\includegraphics[scale=0.40,angle=00]{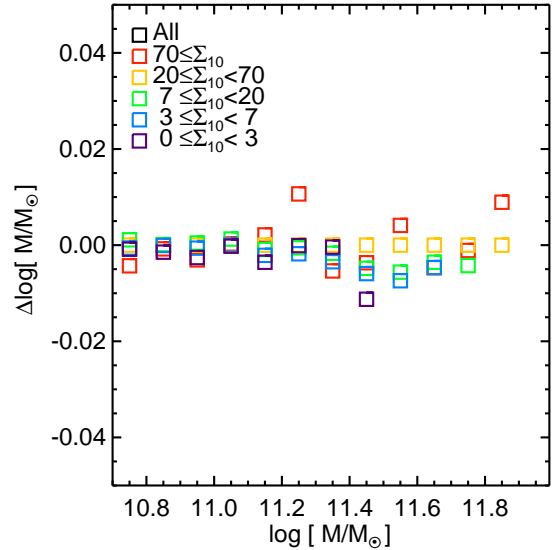}
\centering
	\caption{Difference of the average mass between each stellar mass bin and the stellar mass bin in the environment of $20\leq\Sigma_{10}<70$ in the same mass range.
		\label{massdiff_fig}}
\end{figure}

 We show the mass--size relation of the mass-matched sample in Figure \ref{msmatched_fig}. In this figure, we matched the mass distribution of each mass bin with that of the counterpart in the environment of $20\leq\Sigma_{10}<70$. The use of the mass-matched sample does not change our result. This is related to the very small difference of the average mass between each stellar mass bin in high density and low density in the same mass range, which is measured to be smaller than $\sim0.01$ dex, as shown in Figure \ref{massdiff_fig}. This small mass difference corresponds to a size difference smaller than $\sim0.01$ dex by Equations \ref{eq4} and \ref{eq5}. 
\\

\end{document}